%% file: main.tex
\PassOptionsToPackage{ruled,vlined,algo2e}{algorithm2e}
\documentclass[final,12pt]{colt2026} %

\usepackage{algorithm2e}
\SetAlgoVlined
\SetAlCapHSkip{0pt}
\DontPrintSemicolon
\SetKwFor{For}{for}{do}{}
\SetKwFor{While}{while}{do}{}
\SetKwFor{ForEach}{foreach}{do}{}

\usepackage{booktabs}
\usepackage{thm-restate}
\newtheorem{lemma}[theorem]{Lemma}
\usepackage{booktabs}
\usepackage{pgfplots}
\usepackage{framed}
\usepackage{cancel}
\usepackage{multirow}
\usepackage{enumitem}
\usepackage{bbm}
\usepackage[most]{tcolorbox}

\newcommand{\tv}{\mathrm{d_{TV}}}
\newcommand{\UST}{\mathrm{UST}}
\newcommand{\LFHT}{\mathrm{LFHT}}
\newcommand{\du}{\ensuremath{d_\mathrm{un}}}
\newcommand{\dk}{\ensuremath{d_\mathrm{kn}}}
\newcommand{\eswt}{\ensuremath{\mathtt{ESW\text{-}Tester}}}
\newcommand{\mlfht}{\ensuremath{\mathtt{MultiLFHT}}}

\newcommand{\scheffe}{\ensuremath{\mathtt{Scheff\acute{e}}}}
\setlist[itemize]{topsep=2pt, partopsep=0pt}
\title[A Distribution Testing Approach to Clustering Distributions]{A Distribution Testing Approach to Clustering Distributions}
\usepackage{times}

\coltauthor{%
\Name{Gunjan Kumar} \Email{gunjan@cse.iitk.ac.in}\\
\addr IIT Kanpur
 \AND
 \Name{Yash Pote} 
 \Email{yashppote@gmail.com}\\
  \addr National University of Singapore
 \AND
 \Name{Jonathan Scarlett}\Email{scarlett@comp.nus.edu.sg}\\
 \addr National University of Singapore%
}

\begin{document}

\maketitle

\begin{abstract}
	We study the following distribution clustering problem: Given a hidden partition of $k$ distributions into $2$ groups, such that the distributions within each group are the same, and the distributions associated with the clusters are pairwise $\varepsilon$-far in total variation, the goal is to recover the partition.
	We establish upper and lower bounds on the sample complexity for two fundamental cases: (1) when one of the cluster's distributions is known, and (2) when both are unknown.
	Our upper and lower bounds characterize the sample complexity's dependence on the domain size $n$, number of distributions $k$, size $r$ of one of the clusters, and distance $\varepsilon$. In particular, we achieve tightness with respect to $(n,k,r,\varepsilon)$ (up to an $O(\log k)$ factor) for all regimes.  In addition, we show that this result extends to the case of $d$-clustering for any constant number of clusters $d$.
\end{abstract}


\input{intro}
\input{preliminaries}
\input{upperbounds}

\input{lowerbounds}

\input{discussion}
\input{conclusion}
\acks{This work is supported by the Singapore National Research Foundation (NRF) under its AI Visiting Professorship programme.}
\bibliography{biblio}
\input{supplementary}

\end{document}

%% file: intro.tex
\section{Introduction}\label{sec:intro}
Imagine that there has been a manufacturing defect at the dice factory, and an unknown number of dice are created with a common defect. Unfortunately, all the dice look identical, so the only way to find the fair dice is to roll them and observe the outcomes. How many rolls are needed to determine the set of fair dice?

This is a distribution clustering problem, and is formalised as follows: We are given sample access to distributions $D_1,D_2,\ldots D_k$, and a parameter $\varepsilon$. It is known that there is a hidden partition   $\{\mathcal{I}, [k] \setminus \mathcal{I}\}$ of $[k]$ into two clusters such that all distributions within each cluster are identical, and distributions in different clusters are $\varepsilon$-far in Total Variation ($\tv$), and the goal is to find the hidden partition. We are interested in the sample complexity, i.e., the number of samples required to correctly determine the partition with probability at least $\frac{2}{3}$. This probability can be ``boosted'' to $1-\delta$ by repeating the algorithm $O\big(\log\frac{1}{\delta}\big)$ times.

Broadly speaking, clustering is a fundamental problem in unsupervised learning, where the objective is to group together objects that are similar to each other while separating those that are dissimilar. Clustering of \textit{distributions} has been studied in various forms in previous work, e.g., \citep{ZCY22,YZT24,YHTS25}, as we further discuss in Section~\ref{sec:related}.  Recent advances on this problem, particularly in the literature on bandit algorithms, have focused on designing clustering algorithms with strong asymptotic guarantees. These methods operate in the fixed-confidence setting, where the goal is to output the partition with error at most $\delta$. Certain optimality results are established in the limit $\delta \rightarrow 0$. However, these asymptotic analyses do not provide results for the practical scenario where $\delta$ is constant.  In particular, the precise dependence on important parameters, such as the domain size $n$, is hidden in lower-order terms in these analyses, and remains to be fully characterized.

In contrast, our approach builds on advances in distribution testing (surveyed by~\cite{C20,C22}), where the focus is on finite sample complexity for testing properties of distributions. The goal in distribution testing is to determine whether a given distribution possesses a certain property of interest (such as uniformity) or is far from having it. Over the past decade, this area has seen significant progress, with novel algorithms developed to test various distributional properties, including uniformity, identity, and equivalence. Our work takes a distribution testing approach to clustering distributions, where we assume that the distributions in different clusters are `far' (in total variation distance) and seek to determine the optimal number of required samples  to attain a given constant error probability.

A naive approach for our setting is to fix one distribution, say $D_1$, and then test whether each of the remaining distributions $D_2, \ldots, D_k$ is identical to $D_1$. 
Depending on whether $D_1$ is known or unknown, this can be done using identity or equivalence testing algorithms.
This approach requires $kf(n)$ samples, where $f(n)$ denotes the sample complexity of testing one distribution (see Table~\ref{tab:dist_testing}). 

Is this  dependence unavoidable? Or can the whole task be accomplished with fewer than $k f(n)$ samples?  We show that we can do better than the naive approach by developing a two-stage algorithm that first identifies an exemplar from each cluster and then classifies the rest.

As an initial example, consider the die testing example above, where one kind of distribution is uniform ($U$) and the other is unknown ($P$), and $\varepsilon$-far from uniform. The naive method requires $k \sqrt{n}$ samples, while the optimal sample complexity turns out to be $\sqrt{nk}\max\big(\frac{\sqrt{k}}{r},1\big)$ where $r$ is the number of distributions of the $P$ type. Thus, it is $\sqrt{k}$ times harder to find a loaded die in a bag of fair dice than it is to find a fair die in a bag of loaded dice.  Intuitively, the latter scenario is easier because we can more readily learn information about the unknown distribution $P$.

\input{table}

\subsection{Related Work}\label{sec:related}
\paragraph{A Solved Case: Known Distributions}%
The problem of clustering $k$ distributions into two groups based on known reference distributions $P$ and $Q$ has a simple algorithm known as Scheffe's test~\citep{S47,DL01}. For each distribution $D_i$, the task of deciding whether $D_i=P$ or $D_i=Q$ is achieved by sampling from $D_i$ and estimating the probability of them falling into the Scheffé set $S^+ = \{x:P(x) > Q(x)\}$. This reduces the problem to a hypothesis test on a Bernoulli mean: testing if the mean is $p_0 = Q(S^+)$ or $p_1 = P(S^+)$. Observing that $p_1 - p_0  = \tv(P, Q) = \varepsilon$, and that the complexity to classify a single distribution $D_i$ with error probability $\frac{1}{3k}$ is $O(\frac{\log(k)}{\varepsilon^2})$, we can run $k$ such tests to classify all $k$ distributions with a total error probability of at most $1/3$. The sample complexity is thus $O(\frac{k\log(k)}{\varepsilon^2})$. A near-matching lower bound of $\Omega(\frac{k}{\varepsilon^2})$ is shown by~\cite{MT04}.

\paragraph{Distribution Testing}
Introduced by \citet{GGR98} and \citet{BFRS00}, {distribution testing} is a topic in property testing concerned with deciding whether a distribution has a certain property. For instance, in \textit{uniformity testing}, one must decide if an unknown distribution $Q$ is uniform or $\varepsilon$-far from uniform in total variation distance~\citep{P08}. More generally, \textit{identity testing} involves distinguishing between $\tv(P,Q) = 0$ and $\tv(P,Q) \ge \varepsilon$ for a known distribution $P$ and an unknown distribution $Q$. If both $P$ and $Q$ are unknown, the problem is called \textit{equivalence testing}~\citep{CDVV14}. We use results from these areas as building blocks for our algorithms, and we list them in Table~\ref{tab:dist_testing}.
{\small 
\begin{table}[h!]
    \centering
    \renewcommand{\arraystretch}{1.4} %
    \setlength{\tabcolsep}{12pt} %
    \begin{tabular}{@{}lc@{}}
	        \toprule
	        \textbf{Problem} & \textbf{Complexity} \\
	        \midrule
	        Uniformity/Identity& $ \Theta({\sqrt{n}}/{\varepsilon^2}) $ \\
	        Equivalence  &$
	        \Theta ( \max ( {n^{2/3}}/{\varepsilon^{4/3}}, {\sqrt{n}}/{\varepsilon^2} ) )$
	        \\
	        Learning in TV& $\Theta\left(n/\varepsilon^2\right)$ \\
	        \bottomrule
	    \end{tabular}
	    \caption{A summary of standard distribution testing results, e.g., see \cite{C22}.}
    \label{tab:dist_testing}
\end{table}
}
\citet{G20} showed that every identity testing instance can be converted to a uniformity testing problem over a slightly (6$\times$) larger domain. This connection will be used frequently in our analysis.

\textit{Likelihood-Free Hypothesis Testing} (LFHT) is a related problem that addresses hypothesis testing problems in which the likelihood function is unknown or intractable; this problem is rooted in early works \citep{Z02,G02} and was recently studied in \citep{GHP23,GP24}. The task in LFHT is to determine the optimal trade-off between the number of samples $s_1$ drawn from two candidate distributions, $P$ and $Q$, and the number of samples $s_2$ drawn from a third distribution $D$, to decide whether $D=P$ or $D=Q$.

We note that \textit{clusterability} has been studied under a distribution testing framework by~\cite{LRR13}. A list of distributions is said to be $(c,\beta)$-clusterable if there exists a $c$-clustering where each distribution in a cluster is $\beta$-close in TV distance from every other distribution in the same cluster. A list of distributions is $\varepsilon$-far from being clusterable if the \textit{average} of distances of all distributions from the list from the closest clusterable list is at least $\varepsilon$. Their setting is significantly different from ours, as clusterability is 1) a decision (yes/no) problem, and 2) does not require samples from all distributions, as it is a bulk property. In contrast, clustering requires us to output a full partition, where each distribution is correctly placed.

\paragraph{Bandits and Clustering}
Multi-Armed Bandit techniques have been widely adopted to glean insights from stochastic data in fields ranging from quality control to user behavior analysis.   While the classical goal in this field is maximizing the mean reward, several works have also studied the clustering objective \citep{ZCY22,YZT24,YHTS25}.  Among these, perhaps the most relevant is~\citet{YHTS25}, as they focus on clustering discrete distributions.  However, there are several major differences that make their results incomparable to ours.  Perhaps most importantly, they focus on obtaining tight sample complexity bounds as the error probability $\delta \to 0$ (via the so-called track-and-stop method), but the lower-order asymptotic terms hide the dependence on key parameters such as the domain size and cluster sizes.  We elaborate on this in Appendix \ref{app:bandit}, where we highlight the fundamental differences between fixed-$\delta$ and vanishing-$\delta$, and discuss how the techniques of~\citet{YHTS25} are unsuitable for the former. 

\subsection{Summary of Our Contributions}\label{sec:contributions}
Our main contributions are outlined as follows:
\begin{itemize}
    \item We propose a novel approach to the 2-clustering problem based on distribution testing, providing algorithms that leverage powerful existing learning/testing techniques.
    \item We analyze the sample complexity of our algorithms and provide matching lower bounds (up to logarithmic factors) in all parameter regimes (in terms of $n,r,k$, and $\varepsilon$).
    \item We show that our techniques naturally generalize to the problem of clustering into $d$ groups for any constant $d$, preserving the same dependencies in the sample complexity.
\end{itemize}
A summary of our contributions is presented in Table~\ref{tab:summary_gaps}, with discussion in the next subsection.

\paragraph{Organisation} The remainder of this paper is organized as follows. In Section~\ref{sec:prelims}, we formalize the problem setting. In Section~\ref{sec:upperbounds}, we present our clustering algorithms for the two variants and determine their sample complexity, and provide upper bounds on their sample complexity. We then establish our lower bounds in Section~\ref{sec:lowerbounds}, and outline the extension to more than two clusters in Section \ref{sec:extension_main}.  Finally,  we conclude with directions for future work in Section~\ref{sec:conclusion}.
Deferred proofs and discussions are provided in the appendices.

%% file: table.tex
{\small 
\begin{table*}[t]
    \centering
    \label{tab:summary_gaps}
    \begin{tabular}{lll}
        \toprule
        \textbf{Bound Type} & \textbf{Sample Complexity} & \textbf{Conditions} \\
        \midrule
        \multicolumn{3}{c}{\textbf{Case 1: One distribution unknown}} \\
        \midrule
        \multirow{2}{*}{Upper} & ${O}\left( \frac{k\log(k)\sqrt{n}}{r\varepsilon^2} + \frac{\sqrt{nk\log(k)}}{\varepsilon^2} \right)$ & $n \gtrsim k\log(k)$ \\
        \addlinespace
        & ${O}\left( \frac{k\log(k)\sqrt{n}}{r\varepsilon^2} + \frac{k\log(k)}{\varepsilon^2} \right)$ & $n \lesssim k\log(k)$ \\
        \addlinespace
        \multirow{2}{*}{Lower} & $\Omega \left(\frac{k\sqrt{n}}{r\varepsilon^{2}}\right)$ & $r \leq \frac{k}{60}$ \\
        \addlinespace
        & $\Omega \left(\frac{k+\sqrt{nk}}{\varepsilon^{2}}\right)$ & -- \\
        \midrule
        \multicolumn{3}{c}{\textbf{Case 2: Both distributions unknown}} \\
        \midrule
        \multirow{3}{*}{Upper} & ${O}\left(\max\left(\left(\frac{nk}{r\varepsilon^2}\right)^{2/3}, \frac{k\sqrt{n}}{r\varepsilon^2}\right)\log(k) + \left(\frac{n^2k\log(k)}{\varepsilon^4}\right)^{1/3}\right)$ & $n \gtrsim \frac{k\log(k)}{\varepsilon^4}$ \\
        \addlinespace
        & ${O}\left(\max\left(\left(\frac{nk}{r\varepsilon^2}\right)^{2/3}, \frac{k\sqrt{n}}{r\varepsilon^2}\right)\log(k) + \frac{\sqrt{nk\log(k)}}{\varepsilon^2}\right)$ & $k\log(k) \lesssim n \lesssim \frac{k\log(k)}{\varepsilon^4}$ \\
        \addlinespace
        & ${O}\left(\max\left(\left(\frac{nk}{r\varepsilon^2}\right)^{2/3}, \frac{k\sqrt{n}}{r\varepsilon^2}\right)\log(k) + \frac{k\log(k)}{\varepsilon^2}\right)$ & $n \lesssim k\log(k)$ \\
        \addlinespace
        \multirow{3}{*}{Lower} & $\Omega\left(\max\left(\left(\frac{nk}{r\varepsilon^{2}}\right)^{2/3}, \frac{k\sqrt{n}}{r\varepsilon^{2}}\right)\right)$ &  $ r \leq \frac{k}{60}$ \\
        & $\Omega\left(\left(\frac{n^{2}k}{\varepsilon^{4}} \right)^{1/3}+ \frac{k+\sqrt{nk}}{\varepsilon^{2}}\right)$ & --\\
        \bottomrule
    \end{tabular}
    \caption{A summary of our results. The notation $\lesssim$ and $\gtrsim$ represents inequality to within suitable constant factors.  The upper and lower bounds are shown to coincide to within at most a $\log k$ factor in Appendix \ref{sec:discussion}.} \vspace*{-3ex}
\end{table*}
}

%% file: preliminaries.tex
\subsection{Discussion of Our Results}
\label{sec:prelimdiscussion}

Our results, summarized in Table~\ref{tab:summary_gaps}, establish nearly tight upper and lower bounds on the sample complexity. The gap is at most a $\log(k)$ factor in the upper bounds, which comes from applying a union bound over the $k$ distributions.  See Appendix \ref{sec:discussion} for a detailed description of the tightness of our results. The complexity of our algorithms is the sum of two costs, $T_1 + T_2$, corresponding to two stages: finding an exemplar from each cluster ($T_1$) and classifying the rest ($T_2$). We provide two distinct lower bounds that separately match the complexity of each stage, confirming that this two-stage approach is optimal.

The cost $T_1$ depends on the minority cluster size $r$, with smaller $r$ being hardest (``needle-in-a-haystack''), and this term matches our $r$-dependent lower bounds (Theorem~\ref{thm:indexfindinglb}). In contrast, the classification cost $T_2$ is independent of $r$ and can dominate when finding an exemplar is easy (i.e., for large $r$). This is matched by our $r$-independent lower bounds derived from an LFHT framework (Theorem~\ref{thm:combined-lb-lfht}).  The different regimes for the domain size $n$ arise from using the optimal underlying testing subroutine for the given parameters.

While our upper and lower bounds are written in a different form in Table \ref{tab:summary_gaps}, we will show in Appendix \ref{sec:discussion} that they coincide to within at most a $\log k$ factor in all regimes.

\section{Problem Setup}
\label{sec:prelims}

We consider discrete distributions over the domain $[n] = \{1,2,\ldots n\}$.  We typically adopt the symbol $D$ for a generic distribution over $[n]$, and let $x \sim D$ denote a single sample $x$ drawn from $D$. We define the shorthand $D(S) := \sum_{i \in S} D(i)$ for any set $S$ of elements from the domain of $D$. The total variation distance between two distributions is given as $\tv(D_1, D_2) = \frac{1}{2}\sum_{i \in [n]} |D_1(i) - D_2(i)|$. We write $A \gtrsim B$ to mean $A\geq cB$, and write $A \asymp B$ to mean both $A \lesssim B$ and $B \lesssim A$. 

Our problem setup is summarized as follows.
\begin{framed}
	\noindent\textbf{The Distribution 2-Clustering Problem}
	\begin{description}[leftmargin=1em]
		\item[Input:] Sample access to $k$ distributions $\{D_i\}_{i=1}^k$ over domain $[n]$, parameter $\varepsilon \in (0,1]$.  The distributions are allowed to be queried adaptively (given knowledge of the previous samples).
		\item[Promise:] There exists a hidden partition $\{\mathcal{I}_j\}_{j=1}^{2}$ of $[k]$ satisfying two conditions:
		\begin{itemize}[leftmargin=1.5em]
			\item $\forall{(a, b \in \mathcal{I}_j)}$,  $D_a = D_b$.
			\item $\forall{(a\in \mathcal{I}_1,b \in \mathcal{I}_2)}$, $d_{\mathrm{TV}}(D_a,D_b) \ge \varepsilon$.
		\end{itemize}
		\item[Goal:] With probability at least $2/3$, output the true partition $\{\mathcal{I}_j\}_{j=1}^2$.
		\item[Variants:] We study two settings based on prior knowledge of the two clusters' distributions: {\bf I.}~One is known and the other is unknown, {\bf II.}~Both are unknown.
	\end{description}
\end{framed}
Although we state the problem for a success probability of $\frac{2}{3}$, it is known that repetition and a majority vote can boost the success probability from $\frac{2}{3}$ to $1-\delta$ for any $\delta>0$ at a multiplicative $O(\log(\frac{1}{\delta}))$ cost to the sample complexity. 
Since the output is a partition rather than a binary decision, we describe the boosting explicitly: run the algorithm $m = O(\log(\frac{1}{\delta}))$ times independently, among the $m$ returned partitions we are guaranteed to have $\lceil\frac{m}{2}\rceil+1$ correct clusterings with probability at least $1-\delta$. Since the correct clusterings are all identical, we know the true partition is the unique clustering that appears at least $\lceil\frac{m}{2}\rceil+1$ times.

Given $k$ distributions, the sample complexity of 2-clustering depends on the sizes of the two clusters, which may be known or unknown. We will derive the stronger form of each result accordingly, i.e., our lower bounds are for the case where the cluster size is known, while our upper bounds do not require such knowledge (though we consider the known case as a stepping stone).

For the first variant (\textbf{I}), the relevant factor is the size of the cluster for which the distribution is unknown, and for variant (\textbf{II}), it is the size of the smaller cluster. We denote this factor as $r$, and our bounds will have tight dependences on $r$. Throughout the paper, we assume that the two clusters are non-empty i.e., $r \in [1,k-1]$

\subsection{Adaptivity, Cluster Size Knowledge, and Unknown $\varepsilon$}\label{sec:adaptivity}
Our algorithms are adaptive, i.e., we allow the samples to be drawn sequentially, and the decision of which distribution to sample from at each step can be informed by all past outcomes. However, our algorithms will only require taking samples in two sequential batches if $r$ is known, or $O\big(\log \frac{k}{r}\big)$ sequential batches if $r$ is unknown.  On the other hand, our lower bounds are constructed to hold even for fully adaptive procedures, thus making the near-optimality results stronger.

For clarity of exposition, our algorithms are presented assuming the cluster size $r$ is known.   We describe the variation for unknown $r$ in Appendix~\ref{sec:proofofadaptivity}, and show that it only requires minor modifications and at most a constant factor increase in samples, albeit with more adaptive rounds as mentioned above.   
Similarly, while our main presentation assumes the separation parameter $\varepsilon$ is known, we show in Appendix~\ref{sec:epsilonadaptivity} that this assumption can be relaxed.  Specifically, we provide an adaptive algorithm that does not require knowledge of $\varepsilon$ and achieves the same sample complexity up to $O(\log \log (1/\varepsilon))$ factors and an additive $O\big(\frac{n}{\epsilon^2}\big)$ term (which does not depend on $k$ and is thus non-dominant in several large-$k$ regimes including $k = \Omega(n)$).

%% file: upperbounds.tex
\section{Algorithms for 2-Clustering}\label{sec:upperbounds}
\input{upperbounds/known-unknown}
\input{upperbounds/unknown-unknown}

%% file: upperbounds/known-unknown.tex
\subsection{One Unknown, One Known}\label{sec:1known1unknown}
Here we assume that we are given knowledge of one distribution $Q$, whereas the other distribution $P$ is unknown. Using the Identity-to-Uniformity reduction mentioned in Section~\ref{sec:related}, we can assume without loss of generality that the known distribution ${Q}$ is the uniform distribution $U$, and accordingly, we will henceforth use $U$ instead of ${Q}$.  A formal statement of this reduction is given in Appendix~\ref{sec:idtounif}. Thus, the task in this section is to find the hidden partition $\{\mathcal{I}_{P}, \mathcal{I}_{U}\}$.

\paragraph{Naive Upper Bound}
A naive approach is to test each distribution $D_1, \ldots, D_k$ for uniformity, ensuring a success probability of at least $1-\frac{1}{k}$ for each test. This requires $O\!\left(\frac{\sqrt{n}\log(k)}{\varepsilon^2}\right)$ samples per distribution (see Table~\ref{tab:dist_testing}, with an extra $\log k$ term coming from the union bound), for a total sample complexity of $O\!\left(\frac{k\log(k)\sqrt{n}}{\varepsilon^2}\right)$.

\paragraph{A Better Upper Bound}
Our algorithm improves on the naive approach by proceeding in two stages.  First, we find \textit{one} distribution in $\mathcal{I}_{P}$, which gives us sample access to $P$.  Given this sample access, we can use distribution testing techniques to efficiently classify \emph{all} of the other distributions.

\paragraph{Stage 1: Find One Unknown}
We select a subset of $\lceil \frac{3k}{r} \rceil$ distributions uniformly at random, and test each of them for uniformity. Let $E_1$ be the event that this subset contains no non-uniform distributions. If we upper bound the failure probability by sampling with replacement, then
$
\Pr[E_1]  \le \exp\left(-\frac{r}{k}\cdot \frac{3k}{r}\right)=e^{-3} \le \frac{1}{9}.
$
Note that if $\frac{3k}{r} \geq k$, then the subset contains all $k$ distributions, and hence $\Pr[E_1] = 0$ trivially. To ensure the uniformity tests are reliable, we use enough samples so that the probability of any test failing, $E_2$, is bounded by $\Pr[E_2] \leq \frac{1}{9}$ via a union bound. Assuming the complement event $\overline{E}_1 \wedge \overline{E}_2$ occurs, we successfully identify one distribution of the unknown type.  By Table~\ref{tab:dist_testing} along with an $O\big(\log\frac{k}{r}\big)$ factor for success amplification and the union bound, the sample complexity of this stage is $O\big(\frac{k}{r} \cdot \frac{\sqrt{n}\log(k)}{\varepsilon^2}\big)$. 

\paragraph{Stage 2: Find All Unknowns}
Supposing that we completed Stage 1 successfully, we have identified one of the $k$ distributions as being the unknown distribution $P$. Without loss of generality we will assume $D_1$ is such a distribution. We then attempt to locate all the unknown distributions. To do so, we will use one of two algorithms, $\mlfht$ (Theorem~\ref{thm:multilfht}) and $\eswt$ (Algorithm~\ref{alg:empiricalsubsettest} in Appendix~\ref{sec:esw_analysis}), 
depending on the problem parameters.

We start by presenting the theorem stating the complexity of the $\mlfht$ algorithm.  
\begin{restatable}{theorem}{lfht}\label{thm:multilfht}
Let $P$ and $Q$ be two distributions with $\tv(P, Q) \geq \varepsilon$. Given sample access to $P$ and $Q$, there exists a procedure (which we refer to as $\mlfht$) that, with probability at least $\frac{8}{9}$, correctly classifies distributions $\{D_i\}_{i = 1}^k$ as $P$ or $Q$. The sample complexity is as follows in the $O(\cdot)$ sense:
    $$
    \begin{cases}
        \left(\frac{n^2k\log(k)}{\varepsilon^4}\right)^{1/3} & \text{if } n \gtrsim \frac{k\log(k)}{\varepsilon^4} \\[2ex]
        \frac{\sqrt{nk\log(k)}}{\varepsilon^2} & \text{if } k\log(k) \lesssim n \lesssim \frac{k\log(k)}{\varepsilon^4} \\[2ex]
        \frac{k\log(k)}{\varepsilon^2} & \text{if } n \lesssim k\log(k)
    \end{cases}
    $$
\end{restatable}
\begin{proof}
We establish this result by invoking LFHT multiple times, and using its sample complexity bounds given in Table 2 in~\citep{GHP23}. The full proof is deferred to Appendix~\ref{sec:multilfht}.
\end{proof}

We now present the Empirical Subset Weighting Tester ($\eswt$) (the pseudocode can be found in Appendix~\ref{sec:esw_analysis}). This adapts the test proposed by \cite{ACLS+22}, which was designed to take in two sets of samples of size $s_1$ and $s_2$ from a distribution $D$ to determine whether $D=U$ or $\tv(D,U) \geq \varepsilon$. We adapt the test so that we can draw $s_1$ samples from $P$ and $s_2$ samples from $D$ to distinguish between $D=P$ and $D=U$.

\begin{restatable}{lemma}{eswtcorrectness}\label{lem:esw_tester_correctness}
$\eswt$ takes in sample access to $k$ distributions $\{D_i\}_{i \in [k]}$ over $[n]$, parameter $\varepsilon \gtrsim 1/n^{1/4}$, and sample access to a distribution $P$. It is promised that for an unknown set $\chi \subseteq [k]$, $D_{i} = P$ for $i \in \chi$, and $D_{i}= U$ for $ i \in [k]\setminus\chi$, with $\tv(P,U) \geq \varepsilon$. $\eswt$ outputs $\chi$ with probability at  least $\frac{8}{9}$, and the sample complexity is 
	$$ \begin{cases}
        O\left(\frac{\sqrt{nk\log(k)}}{\varepsilon^2}\right) & \text{if } n \gtrsim \frac{k\log(k)}{\varepsilon^4} \\
        O\left(\frac{k\log(k)}{\varepsilon^4}\right) & \text{if } n \lesssim \frac{k\log(k)}{\varepsilon^4}
    \end{cases} .$$
\end{restatable}

The correctness of the {\eswt} relies on the observation that samples drawn from a non-uniform distribution $P$ tend to collide on ``heavy" elements. Consequently, the set of unique elements $S$ constructed in the first step will carry significantly more probability mass under $P$ than under the uniform distribution $U$. 
    
The detailed analysis, provided in Appendix~\ref{sec:esw_analysis}, formalizes this intuition using two key technical results:
    \begin{itemize}
        \item A separation lemma (Lemma~\ref{lem:eswt}, adapted from~\citep{C22}) which quantifies the gap between the expected mass of $S$ under $P$ versus $U$.
        \item A majorization argument (Lemma~\ref{lem:majorise}) which ensures that the mass of $S$ under $U$, remains bounded with high probability, even when $S$ is generated from $P$.
    \end{itemize}
    
    Given this separation, the algorithm reduces the problem to a simple mean estimation task: for each $D_j$, it estimates the mass $D_j(S)$ using $s_2$ samples. If $D_j = P$, the estimate will be high; if $D_j = U$, it will be low. The sample complexities $s_1$ (for constructing $S$) and $s_2$ (for testing) are optimized to minimize the total cost $s_1 + k s_2$, yielding the two regimes stated in the lemma.

We now put the above findings together and state the resulting theorem, and its associated sample complexity for `Find All'.  The derivation of this sample complexity is given in Appendix~\ref{sec:stage2oneunknown}.
\begin{restatable}{theorem}{oneknown}\label{thm:stage2_complexity_one_unknown}
    Let $P$ be a distribution such that $\tv(P, U) \ge \varepsilon$. There exists an algorithm with sample access to $P$ that, with probability at least $\frac{8}{9}$, correctly classifies all distributions $\{D_i\}_{i = 1}^k$ as $P$ or $U$ . The total sample complexity is as follows:
    \begin{align*}
   \begin{cases}
        O\left(\frac{\sqrt{nk\log(k)}}{\varepsilon^2}\right) & \text{if } n \gtrsim k\log(k) \\[2ex]
        O\left(\frac{k\log(k)}{\varepsilon^2}\right) & \text{if } n \lesssim k\log(k).
    \end{cases}
    \end{align*}
\end{restatable}

\paragraph{Combining the Two Stages} 
The total error of the two stages is bounded by the sum of probabilities of all the bad events: $\Pr[E_1] + \Pr[E_2] + \Pr[\text{Stage 2 Algorithm Fails}] \leq \frac{1}{9} + \frac{1}{9} + \frac{1}{9}  = \frac{1}{3}$. 
Combining the algorithms for the two stages gives us the upper bounds stated in Table~\ref{tab:summary_gaps}, and we state this formally as Theorem~\ref{thm:one_known_upper} in Appendix~\ref{sec:derivation_oneknown}.

%% file: upperbounds/unknown-unknown.tex
\subsection{Both Unknown}\label{sec:bothunknown}

In the case that both distributions are unknown, the naive upper bound, by performing $k-1$ independent equivalence tests (Table~\ref{tab:dist_testing}) is:
$$O\left(k \log(k)\cdot\max\left(\frac{n^{2/3}}{\varepsilon^{4/3}}, \frac{\sqrt{n}}{\varepsilon^2}\right)\right).$$ We improve upon this using similar ideas as our two-stage approach from Section~\ref{sec:1known1unknown}. The first stage involves finding one distribution of each type, and the second stage partially learns both the distributions and then uses that to classify each remaining distribution.
\paragraph{Stage 1: Find One of Each}
Suppose that $D_1$ has the majority distribution $P$; if not, a similar argument applies with the roles of $P$ and $Q$ swapped. Since $r$ of the $k$ distributions are $Q$,  we expect to have $Q$ appear once in a uniformly random sample of $\frac{k}{r}$ distributions. Formally, let $\mathcal{I} \subseteq [k]$ be a uniformly random subset of size $ \lceil \frac{3k}{r} \rceil$, and let $E_1$ be the event that there is at least one $Q$ in $\mathcal{I}$. By a standard tail bound, we upper bound the failure probability by 
$\Pr[\overline{E_1}] \le e^{-3} \le \frac{1}{9}.$

To actually find $Q$, we will adapt the \textit{unequal samples equivalence test} from \citep{DK16}. In our version, we draw $s_1$ samples from $P$(via $D_1$) to create a `sketch' set, which we then use to test the remaining distributions, drawing $s_2$ samples from each. Since the algorithm and proof follow from the ones presented in \citep{DK16}, we defer the details to   Appendix~\ref{sec:unknownunknownupperbound}. The sample complexity of this test is:
\begin{align*}
O\left(\max \left(\left(\frac{nk}{r\varepsilon^2}\right)^{2/3}, \frac{k\sqrt{n}}{r\varepsilon^2}\right)\log(k)\right).
\end{align*}

We choose $s_1$ and $s_2$ such that each test is erroneous with probability at most $\frac{1}{9\cdot \lceil 3k/r\rceil}$, and we denote the event that at least one test fails by $E_2$. Using the union bound over $\lceil\frac{3k}{r}\rceil$ equivalence tests, we obtain $\Pr[E_2] \leq \frac{1}{9}$. We will henceforth assume $\overline{{E}_2}$ occurs.  
\paragraph{Stage 2: Find All}
Assuming success in Stage 1, we have identified two distributions as being $P$ and $Q$. We use the $\mlfht$ algorithm from Theorem~\ref{thm:multilfht} to classify the remaining $k-2$ distributions into their respective clusters with probability at least $\frac{8}{9}$.
\paragraph{Combining the Two Stages} 
The total error probability of the two-stage algorithm bounded by the sum of all failure events: $\Pr[\overline{E_1}] + \Pr[E_2] + \Pr[\text{Stage 2 Fails}] \leq \frac{1}{9} + \frac{1}{9} + \frac{1}{9} = \frac{1}{3}$.
The sum of the costs from Stage 1 and Stage 2 gives us the upper bound as stated in Case 2 of Table~\ref{tab:summary_gaps}. The formal statement of these bounds is given as Theorem~\ref{thm:both_unknown_upper} in Appendix~\ref{sec:derivationbothunknown}.

%% file: lowerbounds.tex
\section{2-Clustering Lower Bounds}\label{sec:lowerbounds}
We will present two lower bounds, one in Section \ref{sec:generalised} and another in Section \ref{sec:LFHTgeneralised}, which will be relevant for different parameter regimes. The lower bound provided in Section \ref{sec:generalised} can be roughly interpreted as a lower bound `against' Stage $1$ of our algorithm, whereas the lower bound presented in Section~\ref{sec:LFHTgeneralised} is a bound `against' Stage $2$. These two lower bounds will imply all of our lower bounds in Table~\ref{tab:summary_gaps}. We reiterate that all of our lower bounds hold even when the cluster sizes ($r$ and $k-r$) are known by the algorithm, and even when the algorithm may be fully adaptive.

In Section~\ref{sec:generalised}, we establish a lower bound of 
$\Omega\!\left(\tfrac{k \sqrt{n}}{r \varepsilon^{2}}\right)$ for the 
\emph{One-Known-One-Unknown} case, and a lower bound of $
\Omega\!\left(\max\!\left\{ \left(\tfrac{nk}{r \varepsilon^{2}}\right)^{2/3} , \tfrac{k \sqrt{n}}{r \varepsilon^{2}}\right\} \right)$
for the \emph{Both-Unknown} case. Both bounds follow from a common framework: 
We reduce \emph{from} uniformity/equivalence testing to the task of determining an index of a distribution belonging to a particular cluster—an easier problem than the full clustering task—thereby inheriting the known lower bounds for $\UST$.

In Section~\ref{sec:LFHTgeneralised}, we establish a lower bound of 
$\Omega\!\left(\tfrac{\sqrt{nk}}{\varepsilon^{2}}\right)$ for the 
\emph{One-Known-One-Unknown} case, and a lower bound of 
$\Omega\!\left(\max\!\left\{ \tfrac{k+\sqrt{nk}}{\varepsilon^{2}}, 
\left(\tfrac{kn^{2}}{\varepsilon^{4}}\right)^{1/3}\right\}\right)$ for the 
\emph{Both-Unknown} case. These bounds also arise from a reduction, but a different one from that of Section~\ref{sec:generalised}. 
Specifically, here we reduce from known lower bound results in likelihood-free hypothesis testing ($\LFHT$), showing that any algorithm for our clustering problem can be converted into an algorithm for $\LFHT$.

\subsection{Clustering Lower Bounds via Unequal Sample Distribution Testing}\label{sec:generalised}

As mentioned before, we prove our first lower bounds by showing lower bounds for an easier task: identifying the index of a \emph{single} distribution from a specific cluster. When one distribution is known (assumed uniform), this reduces to finding the index of any non-uniform distribution. When both distributions are unknown, it reduces to identifying the index of a distribution from the smaller cluster (here we assume $r\leq \frac{k}{60}$, so the clusters are of unequal size).

\begin{restatable}{theorem}{indexfindinglb}\label{thm:indexfindinglb}
Let $k \geq 60$ and $1 \leq r \leq \frac{k}{60}$. Consider a set of $k$ distributions $\{D_i\}_{i \in [k]}$, where for an unknown subset $\mathcal{I} \subseteq [k]$ of size $r$, we have $D_i = P$ for all $i \in \mathcal{I}$ and $D_i = Q$ for all $i \in [k] \setminus \mathcal{I}$, with $\tv(P,Q) \geq \varepsilon$. Any algorithm (possibly adaptive and/or knowing $r$) that returns an index in $\mathcal{I}$ with probability at least $0.9$ requires a number of samples lower bounded as follows:
\begin{enumerate}[itemsep=0pt]
    \item[(a)] \textbf{(One Known)} If $Q$ is known to be uniform, then the required number of samples is $\Omega\!\left(\tfrac{k \sqrt{n}}{r \varepsilon^2}\right)$.
    \item[(b)] \textbf{(Both Unknown)} If both $P$ and $Q$ are unknown, the required number of samples is $\Omega\!\left(\max\!\left\{ \tfrac{k \sqrt{n}}{r\varepsilon^{2}}, \left(\tfrac{nk}{r \varepsilon^{2}}\right)^{2/3} \right\}\right).$
\end{enumerate}
\end{restatable}
These bounds are obtained via reductions from unequal sample testing ($\UST$). Formally, $\UST(P,Q,s_1,s_2)$ is the problem of distinguishing with probability at least $0.9$ between $H_0: P=Q$ and $H_1: \tv(P,Q) \ge \varepsilon$ given $s_1$ samples from $P$ and $s_2$ from $Q$. When $Q$ is known to be the uniform distribution $U$, we denote the problem by $\UST(P,U,s_1,s_2)$. These are the standard problems of equivalence and uniformity testing in the unequal-sample setting~\cite{BV15,C22}. The required sample complexity lower bounds are stated as follows.
\begin{lemma} \label{lem:unequal}
Given $\tv(P,U) \ge \varepsilon$ and $\tv(P,Q) \ge \varepsilon$:
\begin{enumerate}
    \item $\UST(P,U,s_1,s_2)$ requires that:  $s_2 \gtrsim \frac{\sqrt{n}}{\varepsilon^2}$, (Thm. 4 in~\cite{P08}).
    \item $\UST(P,Q,s_1,s_2)$ requires that: $
     s_1s_2^2 \gtrsim (\frac{n}{\varepsilon^2})^2 \quad   s_2 \gtrsim \frac{\sqrt{n}}{\varepsilon^2}$, (Thm. 1 in~\cite{BV15}).
     \end{enumerate}
\end{lemma}
Using this, we prove both parts of Theorem \ref{thm:indexfindinglb} in Appendix~\ref{sec:testinglb}. To give an overview of our reduction, here we sketch the argument for the \emph{One-Known-One-Unknown} case with $r=1$, i.e., exactly one index corresponds to an unknown distribution that is far from uniform and the remaining $k-1$ distributions are uniform. First, consider the null instance where all distributions are uniform. By an averaging argument, there must exist an index $j$ whose corresponding distribution  is sampled from in expectation at most $O(T/k)$ times and is output with probability at most $O(1/k)$. 

Now set the distribution $D_j$ at index $j$ to $P$ ($P$ is an instance in $\UST(P,U,s_1,s_2)$, so $P$ can either be $U$ or far from $U$). Correctness requires the algorithm to output $j$ if $P$ is non-uniform with high probability (whereas in the null case $j$ is output with probability at most $O(1/k)$). Moreover, we will abort the algorithm as soon as $D_j$ is sampled $C T/k$ times for some suitably-chosen constant $C>1$. 
Note that if $P$ is far from uniform, then with high probability, either $j$ is output or $D_j$ gets sampled $CT/k$ times. Moreover, if $P$ is uniform, then with high probability, index $j$ is not output and $D_j$ gets sampled fewer than $CT/k$ times. 
Finally, combining that the distribution $D_j$ is always sampled at most $CT/k$ times with the first part of Lemma~\ref{lem:unequal}, we must have $CT/k  \gtrsim \frac{\sqrt{n}}{\varepsilon^2}$ implying $T \gtrsim k \sqrt{n}/\varepsilon^2$.

\subsection{ Clustering Lower Bounds via LFHT Lower Bounds}\label{sec:LFHTgeneralised}

Our second set of lower bounds is stated as follows.

\begin{restatable}{theorem}{lfhtlb}\label{thm:combined-lb-lfht}
    Let $n \geq 1$, $k\geq 100$, and $\varepsilon \in (0,1]$. Consider a set of $k$ distributions $\{D_i\}_{i \in [k]}$ where for an unknown subset $\mathcal{I} \subseteq [k]$ of size $r$, we have $D_i = P$ for all $i \in \mathcal{I}$ and $D_i=Q$ for all  $i \in [k] \setminus \mathcal{I}$ with $\tv(P,Q) \geq \varepsilon$. Any algorithm (possibly adaptive and/or knowing $r$) that returns the partition $\{\mathcal{I}, [k]\setminus \mathcal{I}\}$ with probability at least $0.9$ must have a total sample complexity of:
    \begin{itemize}
        \item[(a)] $\Omega\!\left(\tfrac{k+\sqrt{nk}}{\varepsilon^2}\right)$, when one of the underlying distributions is uniform.
        \item[(b)] $\Omega\!\left(
            \left(\tfrac{kn^2}{\varepsilon^{4}} \right)^{1/3} +\tfrac{k+\sqrt{nk}}{\varepsilon^{2}} 
      \right)
       $ when both underlying distributions are unknown.
    \end{itemize}
\end{restatable}

We briefly outline the proof here, and provide the details in Appendix~\ref{sec:proofoflfhtlb}.  
As mentioned earlier, these bounds are obtained via reductions from \emph{likelihood-free hypothesis testing} ($\LFHT$).
Recall that $\LFHT(P,Q,s_1,s_2)$ is the problem of distinguishing, with probability at least $0.9$, between $H_0 = P^{\otimes s_1} \otimes Q^{\otimes s_1} \otimes P^{\otimes s_2}$ and  
$H_1 = P^{\otimes s_1} \otimes Q^{\otimes s_1} \otimes Q^{\otimes s_2}$ 
where $P,Q$ are distributions with $\tv(P,Q) \ge \varepsilon$. Equivalently, we get $s_1$ samples from each of $P$ and $Q$, and $s_2$ samples from a distribution $D \in \{P,Q\}$, and must decide (w.p.\ $\ge 0.9$) whether $D=P$ or $D=Q$.  When $Q$ is known to be the uniform distribution $U$, $\LFHT(P,Q,s_1,s_2)$ reduces to distinguishing $H_0 = Q^{\otimes s_1} \otimes U^{\otimes s_2}$ from $H_1 = Q^{\otimes s_1} \otimes Q^{\otimes s_2}$; we denote this case by $\LFHT(P,U,s_1,s_2)$.  

\cite{GP24} present tight bounds for these two problems, and we state the relevant lower bounds in the following lemma, again using the notation $A \gtrsim B$ for $A\geq cB$.
\begin{lemma} \label{lem:lfht}
Given $\tv(P,U) \ge \varepsilon$ and $\tv(P,Q) \ge \varepsilon$, and assuming the restriction $s_1 \gtrsim s_2$:
\begin{enumerate}
    \item $\LFHT(P,U,s_1,s_2)$ requires the following constraints: $s_1 \gtrsim \frac{\sqrt{n}}{\varepsilon^2}$, $s_1s_2 \gtrsim \frac{n}{\varepsilon^4},$ and
    $s_2 \gtrsim \frac{1}{\varepsilon^2}$.   
    \item $\LFHT(P,Q,s_1,s_2)$ requires 
\begin{align*}
    \begin{cases}
             s_1^2s_2 \gtrsim (\frac{n}{\varepsilon^2})^2 \quad  s_1s_2 \gtrsim \frac{n}{\varepsilon^4} \quad s_2 \gtrsim \frac{1}{\varepsilon^2}  & \text{if } n \gtrsim \frac{1}{\varepsilon^4}\\
         s_1 \gtrsim \frac{\sqrt{n}}{\varepsilon^2}\quad  s_1s_2 \gtrsim \frac{n}{\varepsilon^4} \quad
    s_2 \gtrsim \frac{1}{\varepsilon^2} & \text{if } n \lesssim \frac{1}{\varepsilon^4}
    \end{cases}
\end{align*}
\end{enumerate}
\end{lemma}
\begin{remark}
    The first bound is inferred from the  construction of $P$ and $Q$ in Section B.3 of~\cite{GP24}, where $Q$ is taken to be the uniform distribution, and the second is from Table 2 of \cite{GHP23}, where the symbols $n,m,$ and $k$ map to $s_1,s_2$ and $n$ in our paper. 
    We state the inequalities in Lemma~\ref{lem:lfht} under the additional constraint $s_1 \gtrsim s_2$, as this will turn out to be automatically satisfied in our reduction argument.
\end{remark}
Our proof of Theorem~\ref{thm:combined-lb-lfht} is based on a switching argument. Note that the algorithm returns a partition of size $(r,k-r)$. Let $\mathcal{I}$ denote the subset of size $k-r$.

We first define a null instance and identify an index $j$ whose corresponding distribution is sampled only a small number of times (in expectation) and also has a small probability of appearing in the output set $\mathcal{I}$. We then replace the distribution at this index with $D$, from an $\LFHT(P,Q,s_1,s_2)$ instance, where $D$ is either $P$ or $Q$.

Depending on whether $D=P$ or $D=Q$, the algorithm behaviour differs, either in the number of times the distribution at index $j$ is sampled from, or in whether $j$ is included in the output set $\mathcal{I}$. This difference is used  to detect whether $D$ is $P$ or $Q$.
 Moreover, distribution at index $j$ (note that this is unknown $D$) is sampled from at most $80T/k$ times, while every other distribution is sampled from at most a total of $T$ times. Substituting $s_1 = T$ (the total number of samples drawn from $P$ and $Q$)  and $s_2 = 80T/k$ (the total number of samples drawn  from unknown $D$) into Lemma~\ref{lem:lfht} yields the desired lower bound on the sample complexity $T$.

%% file: discussion.tex
\section{Extension to $d$-Clustering}\label{sec:extension_main}

Our results naturally generalize to the problem of clustering into $d$ groups for any constant $d > 2$. We consider a setting with $d$ total clusters, comprising $\dk$ known centers and $\du$ unknown centers (where $d = \du + \dk$). 

The algorithmic strategy generalizes our two-stage approach: first, finding $\du$ unknown exemplars using a ``coupon collector'' strategy combined with identity and unequal-samples equivalence testing; and second, running a tournament of pairwise tests to classify the remaining distributions.

\begin{theorem}[Sample Complexity of $d$-Clustering]\label{thm:d_clustering_summary}
    There exists an algorithm that solves the $d$-clustering problem with probability at least $2/3$, and such that for $\du = 1$ (resp., $\du \ge 2$), the sample complexity is as given in Case 1 (resp., Case 2) of Table~\ref{tab:summary_gaps}, up to constant factors that may depend on $d$, where $r$ is the size of the smallest cluster among those with unknown distributions. 
\end{theorem}

We now argue that this sample complexity is tight (up to logarithmic factors).  This is because the $2$-clustering problem is a special case of $d$-clustering, so the lower bounds established in Section~\ref{sec:lowerbounds} apply immediately. Consequently, for any constant $d$, the sample complexity of $d$-clustering is governed by the same tight dependencies on $n, k, r,$ and $\varepsilon$ as the 2-clustering case; recall that this tightness is verified for 2-clustering in Appendix~\ref{sec:discussion}.

The formal algorithms and proofs for this extension are provided in Appendix~\ref{sec:extension}.  

%% file: conclusion.tex
\section{Conclusion and Future Work}\label{sec:conclusion}
We have provided upper and lower bounds for $d$-clustering of distributions, with near-matching scaling in all parameter regimes. Among other things, our results imply the following non-obvious insight: \textit{it is $\sqrt{k}$ times harder to find a loaded die in a bag of fair dice than it is to find a fair die in a bag of loaded dice.}

We conclude by listing some open problems that arise naturally: 

\begin{itemize}
    \item \textbf{Dependence on the failure probability.} As discussed in Appendix~\ref{app:bandit}, it remains open to establish the optimal sample complexity when the dependence on the failure probability $\delta$ is explicitly incorporated. 

    \item \textbf{Optimal bounds for $d$-clustering for $d = \omega(1)$.} While we have provided an optimality result for $d$-clustering with $d = O(1)$, determining the optimal dependence on the number of clusters $d$ remains an open question for $d \in \omega(1)$.

    \item \textbf{Adaptivity gap.} Our algorithm crucially relies on adaptive sampling in order to find a ``representative'' from one cluster (or both) and then use it to classify the remaining distributions. It would be of interest to determine whether adaptivity is essential, i.e., whether non-adaptive sampling requires strictly more samples, thus creating an ``adaptivity gap''.
\end{itemize}

%% file: supplementary.tex
    \appendix
    \section{Comparison to Bandit Clustering Results} \label{app:bandit}
    
    In this section, we provide a more detailed discussion of how our results differ from those of \citep{YHTS25}, which gave a general matching/clustering framework that includes 2-clustering as a special case.
    
    We first highlight that \citep{YHTS25} provides tight results in the asymptotic limit $\delta \to 0$ (with all other parameters held fixed), which is fundamentally different from our attention on constant error probability.  To appreciate this distinction, we note that in the simpler uniformity testing problem with domain size $n$ and TV distance $\varepsilon$, the optimal scaling in general is $\Theta\big( \frac{1}{\varepsilon^2} \big( \sqrt{n \log \frac{1}{\delta}} + \log\frac{1}{\delta} \big) \big)$ \cite[Eq.~(2.50)]{C22}.  Thus, in the limit $\delta \to 0$, the leading term is $\frac{1}{\varepsilon^2} \log\frac{1}{\delta}$.  However, taking this limit neglects the important $\sqrt{n}$ dependence in the general bound.  For instance, one could have derived an upper bound with $\sqrt{n \log \frac{1}{\delta}}$ replaced by $n\sqrt{\log \frac{1}{\delta}}$ or even $2^n\sqrt{\log \frac{1}{\delta}}$, and it would still have been asymptotically optimal in the limit $\delta \to 0$, yet highly suboptimal for constant $\delta$.
    
    Let $\mathbf{D} = (D_1,\dotsc,D_k)$ be the $k$ distributions in the 2-clustering problem.  
    The (expected) sample complexity derived in \citep[Thm.~2]{YHTS25} is instance dependent (i.e., depends on $\mathbf{D}$), and takes the following form as $\delta \to 0$:
    \begin{gather}
        \mathbb{E}[ \#\text{samples}] = \frac{1}{T^*(\mathbf{D})} \log\frac{1}{\delta} \cdot (1+o(1)), \\
        T^*(\mathbf{D}) = \sup_w \inf_{\mathbf{D}' \in {\rm Alt}(\mathbf{D})} \sum_{i=1}^k w_i d_{\rm KL}(D_i, D'_i), \label{eq:T*}
    \end{gather}
    where $w \in [0,1]^k$ is optimized over the probability simplex, and ${\rm Alt}(\mathbf{D})$ is the set of alternative instances whose correct answer is different from $\mathbf{D}$.  We focus here on the case of known cluster sizes, meaning that both $\mathbf{D}$ and $\mathbf{D}'$ have $r$ clusters of one type and $k-r$ of the other. 
    
    We can simplify $T^*(\mathbf{D})$ for 2-clustering by noting the following:
    \begin{itemize}
        \item By the non-negativity of KL divergence, the minimizing $\mathbf{D}'$ swaps just two distributions (one from each cluster) with one another. 
        \item The maximizing $w$ places equal probability within the $r$ distributions in one cluster, and also within the $k-r$ distributions in the other cluster.  This is because, with the intra-cluster distributions being identical, the $\inf_{\mathbf{D}'}$ will be achieved by picking the minimum-weight element of each cluster; then, using uniform weights ensures that each cluster's minimum weight is as high as possible.
    \end{itemize}
    Accordingly, we fix $\lambda \in [0,1]$ and assign weights $\frac{\lambda}{r}$ and $\frac{1-\lambda}{k-r}$ within the size-$r$ and size-$(k-r)$ clusters respectively.  Then, the quantity $\sum_{i=1}^k w_i d_{\rm KL}(D_i, D'_i)$ in \eqref{eq:T*} becomes $\frac{\lambda}{r} d_{\rm KL}(P,Q) + \frac{1-\lambda}{k-r} d_{\rm KL}(Q,P)$ with $P$ and $Q$ being the two clusters' distributions.  Applying Pinsker's inequality and $d_{\rm TV}(P,Q) \ge \varepsilon$ gives $d_{\rm KL}(\cdot,\cdot) \ge 2\varepsilon^2$, so the preceding expression is lower bounded by $2\varepsilon^2 \big( \frac{\lambda}{r} + \frac{1-\lambda}{k-r}\big)$.  By handling the cases $r < \frac{k}{2}$ and $r > \frac{k}{2}$ separately, we readily obtain that this behaves as $\Theta\big( \frac{\varepsilon^2}{k} \big)$ upon maximizing over $\lambda$ (in accordance with the $\sup_w$ operation), and it follows that the sample complexity is upper bounded by $O\big( \frac{k}{\varepsilon^2} \log\frac{1}{\delta}\big)$ as $\delta \to 0$.
    
    In analogy with the earlier discussion on uniformity testing, this asymptotic analysis hides important dependencies -- the result shows no dependence on the domain size $n$ or cluster size $r$, because any such dependence is hidden in lower-order terms in $\delta$.  Moreover, an inspection of the proof in \citep{YHTS25} reveals that their analysis is unsuitable for optimizing such dependencies; for instance, the proof of Lemma 14 therein is based on showing that a quantity of the form $A e^{-B \log^2 t}$ approaches zero for all $t \ge \Omega\big( \log\frac{1}{\delta} \big)$, where $A$ and $B$ are complicated functions of the problem parameters.  We also note that a linear dependence on the domain size (which can often be avoided via distributed testing methods; see Table \ref{tab:dist_testing}) appears much earlier in their analysis, e.g., see Eq.~(66) therein.
    
    On the other hand, it is worth highlighting that the above-mentioned results from \citep{YHTS25} indicate that our approach of amplifying a $2/3$-success guarantee to $1-\delta$ (with $O\big(\log\frac{1}{\delta}\big)$ multiplicative overhead) is \emph{not} optimal in general.  Thus, it remains an open problem to understand the joint dependencies on all parameters \emph{including $\delta$}, thus interpolating between our constant-$\delta$ results and the existing (very) small-$\delta$ results.
    
    \section{Identity To Uniformity Reduction}\label{sec:idtounif}
    The Identity To Uniformity reduction transforms the problem of testing identity to a known distribution $D$ into the problem of testing uniformity \cite{G20}. The reduction is a randomised procedure that maps the original domain $[n]$ to a domain $[n']$ of size $n' = 6n$. This mapping constructs a transformed distribution $P'$ from the unknown distribution $P$ such that the reference distribution $D$ maps exactly to the uniform distribution $U_{6n}$.
    
    \begin{theorem}[Theorem 1 from \citep{G20}]\label{thm:idtounif}
    For every distribution $P$ over $[n]$ and every $\varepsilon > 0$, determining $\tv(P,Q) =0$ vs. $\tv(P,Q)\geq \varepsilon$ reduces to testing $\tv(P',U_{6n}) = 0$ vs. $\tv(P',U_{6n}) \geq \varepsilon/3$, where $U_m$ denotes the uniform distribution over $[m]$. Furthermore, the same reduction $F$ can be used for all $\varepsilon > 0$. 
    \end{theorem}
    Theorem~\ref{thm:idtounif} guarantees that this transformation preserves the distance separation required for testing:
    \begin{itemize}
        \item If $P = D$, then $P' = U_{6n}$.
        \item If $\tv(P, D) \geq \varepsilon$, then $\tv(P', U_{6n}) \geq \varepsilon/3$.
    \end{itemize}
    
    Consequently, any algorithm that tests uniformity over $[6n]$ with distance parameter $\varepsilon/3$ also tests identity to $D$ over $[n]$ with parameter $\varepsilon$. This allows the use of standard uniformity testing bounds for the general identity testing problem.
    
    \section{Derivation of Algorithmic Upper Bounds}
        This section provides a detailed derivation of the sample complexity upper bounds for the algorithms presented in Section~\ref{sec:upperbounds}. We show how the complexities of the two stages combine under different parameter regimes.
    
\subsection{Proof of Theorem~\ref{thm:multilfht} (Sample Complexity of $\mlfht$)}\label{sec:multilfht}

The $\mlfht$ procedure classifies a set of $k$ distributions $\{D_i\}_{i=1}^k$, each promised to be either $P$ or $Q$, by applying a Likelihood-Free Hypothesis Testing (LFHT) subroutine to each one. 
To achieve the target total error probability of at most $\frac{1}{9}$ even after a union bound, the error tolerance for each of the $k$ individual classification tests is set to $\delta=\frac{1}{9k}$. The procedure involves drawing a number of samples (denoted $s_1$)  from the exemplar distributions $P$ and $Q$, and a number of samples (denoted $s_2$) from each distribution $D_i$ that is being classified. We state the sample complexity as follows:

\lfht*
\begin{proof}
	The proof is based on choosing parameters to minimize the total ``cost'' (number of samples) $C = s_1 + ks_2$ subject to suitable constraints in various scaling regimes. To achieve a total error probability of at most $\frac{1}{9}$, we set the per-distribution classification error to $\delta = \frac{1}{9k}$ and apply a union bound. The necessary sample sizes must satisfy several constraints given in Table 2 of \citep{GHP23}. To state the bounds in the language of our paper, we replace the symbols $(n,m,k)$ in their paper with $(s_1,s_2,n)$.
	
	The first constraint is common for all parameter regimes:
	\begin{align}
		s_2 \gtrsim \frac{\log(k)}{\varepsilon^2} \label{eq:s2}
	\end{align}
	The other constraints depend on the domain size $n$:
	\begin{align}
		(\text{if } n \gtrsim \frac{\log(k)}{\varepsilon^4})    &\qquad    s_1^2s_2 \gtrsim \frac{n^2\log(k)}{\varepsilon^4} \quad \text{and} \quad  s_1s_2 \gtrsim \frac{n\log(k)}{\varepsilon^4}\label{eq:nlarge} \\
		(\text{if } n \lesssim \frac{\log(k)}{\varepsilon^4})    &\qquad    s_1s_2 \gtrsim \frac{ (\sqrt{n\log(k)}+\log(k))^2}{\varepsilon^4} \quad \text{and} \quad s_1 \gtrsim \frac{\sqrt{n\log(k)}+\log(k)}{\varepsilon^2}
		\label{eq:nsmall}
	\end{align}
	The optimal strategy for choosing $s_1$ and $s_2$ changes depending on which constraint is dominant, leading to three distinct regimes for the total complexity. 
	\paragraph{Regime 1: $n \gtrsim \frac{k\log(k)}{\varepsilon^4}$}
	In this regime, the $s_1^2s_2$ constraint from (\ref{eq:nlarge}) is the more restrictive of the two. The cost $C = s_1 + ks_2$ is minimized by balancing the two terms, leading to the choice $s_1 \asymp \left(\frac{n^2k\log(k)}{\varepsilon^4}\right)^{1/3}$ and $s_2 \asymp  \left(\frac{n^2\log(k)}{k^2\varepsilon^4}\right)^{1/3}$. The total cost is $O\left(\left(\frac{n^2k\log(k)}{\varepsilon^4}\right)^{1/3}\right)$, and all constraints are satisfied, as we verify as follows:
	\begin{itemize}
		\item The $s_1^2s_2$ constraint is met by construction.
		\item The $s_1s_2$ constraint requires $\left(\frac{n^4\log^2(k)}{k\varepsilon^8}\right)^{1/3} \gtrsim \frac{n\log(k)}{\varepsilon^4}$, which simplifies to $n \gtrsim \frac{k\log(k)}{\varepsilon^4}$, which in turn holds by the definition of the regime.
		\item The $s_2$ constraint requires $\left(\frac{n^2\log(k)}{k^2\varepsilon^4}\right)^{1/3} \gtrsim \frac{\log(k)}{\varepsilon^2}$, which simplifies to $n \gtrsim \frac{k\log(k)}{\varepsilon}$ and thus also holds.
	\end{itemize}
	\paragraph{Regime 2: $k\log(k) \lesssim n \lesssim \frac{k\log(k)}{\varepsilon^4}$}
	Since $n \gtrsim k\log(k)$, we have $\sqrt{n\log(k)} \gtrsim \log(k)$, and thus the $s_1s_2$ constraint from (\ref{eq:nsmall}) becomes $s_1s_2 \gtrsim \frac{n\log(k)}{\varepsilon^4}$, matching (\ref{eq:nlarge}). 
	Our problem is to minimize $C = s_1 + ks_2$ subject to all the relevant constraints. We minimize the cost by balancing the terms, $s_1 \asymp ks_2$.
	\begin{itemize}
		\item Substituting $s_1 \asymp ks_2$ into the $s_1s_2$ constraint gives
		$$(ks_2)s_2 \gtrsim \frac{n\log(k)}{\varepsilon^4} \iff s_2^2 \gtrsim \frac{n\log(k)}{k\varepsilon^4} \implies \text{Can set } s_2 \asymp \frac{1}{\varepsilon^2}\sqrt{\frac{n\log(k)}{k}}.$$
		\item This determines $s_1$ as follows:
		$$s_1 \asymp ks_2 \asymp \frac{\sqrt{nk\log(k)}}{\varepsilon^2}.$$ 
	\end{itemize}
	We verify this solution against all constraints:
	\begin{itemize}
		\item The $s_1$ constraint from (\ref{eq:nsmall}) is satisfied as $s_1 \asymp \frac{\sqrt{nk\log(k)}}{\varepsilon^2} \gtrsim \frac{\sqrt{n \log(k)} + \log(k)}{\varepsilon^2}$. 
		\item The $s_1s_2$ constraint from (\ref{eq:nlarge}) is met by construction: $s_1s_2 \asymp \left(\frac{\sqrt{nk\log(k)}}{\varepsilon^2}\right) \left(\frac{1}{\varepsilon^2}\sqrt{\frac{n\log(k)}{k}}\right) = \frac{n\log(k)}{\varepsilon^4}$.
		\item The $s_2$ constraint from (\ref{eq:s2}) requires $s_2 \gtrsim \frac{\log(k)}{\varepsilon^2}$. Our choice $s_2 \asymp \frac{1}{\varepsilon^2}\sqrt{\frac{n\log(k)}{k}}$ satisfies this, since:
		$$\frac{1}{\varepsilon^2}\sqrt{\frac{n\log(k)}{k}} \gtrsim \frac{\log(k)}{\varepsilon^2} \iff \sqrt{\frac{n}{k\log(k)}} \gtrsim 1 \iff n \gtrsim k\log(k),$$
		which holds by the definition of the regime.
		\item The $s_1^2s_2$ constraint from (\ref{eq:nlarge}) requires $s_1^2s_2 \gtrsim \frac{n^2\log(k)}{\varepsilon^4}$. We have
		$$s_1^2s_2 \asymp \left(\frac{\sqrt{nk\log(k)}}{\varepsilon^2}\right)^2 \left(\frac{1}{\varepsilon^2}\sqrt{\frac{n\log(k)}{k}}\right) = \left(\frac{nk\log(k)}{\varepsilon^4}\right) \left(\frac{\sqrt{n\log(k)}}{\sqrt{k}\varepsilon^2}\right) = \frac{n\sqrt{nk}\log(k)^{3/2}}{\varepsilon^6}.$$
		The constraint is thus satisfied if $\frac{n\sqrt{nk}\log(k)^{3/2}}{\varepsilon^6} \gtrsim \frac{n^2\log(k)}{\varepsilon^4}$, which simplifies to $\frac{\sqrt{k\log(k)}}{\varepsilon^2} \gtrsim \sqrt{n}$ and thus $\frac{k\log(k)}{\varepsilon^4} \gtrsim n$. This holds by the definition of the regime.
	\end{itemize}
	Thus, all constraints are satisfied, and the total cost is $O\left(\frac{\sqrt{nk\log(k)}}{\varepsilon^2}\right)$.
	
	\paragraph{Regime 3: $n \lesssim k\log(k)$}
	In this range, we choose $s_2 \asymp \frac{\log(k)}{\varepsilon^2}$ and $s_1 \asymp \frac{k\log(k)}{\varepsilon^2}$. The total cost $C = s_1 + ks_2$  is thus $O\left(\frac{k\log(k)}{\varepsilon^2}\right)$. We verify the constraints as follows:
	\begin{itemize}
		\item The $s_2$ constraint from (\ref{eq:s2}) is met by construction, as $s_2 \asymp \frac{\log(k)}{\varepsilon^2}$.
		\item The $s_1s_2$ constraints from (\ref{eq:nlarge}) and (\ref{eq:nsmall}) are met, as $s_1s_2 \asymp \left( \frac{k\log(k)}{\varepsilon^2} \right) \left( \frac{\log(k)}{\varepsilon^2} \right) \gtrsim \frac{k\log(k)^2}{\varepsilon^4} \gtrsim \frac{{n\log(k)}}{\varepsilon^4}$ by the definition of the regime.
        
		\item The $s_1^2s_2$ constraint from (\ref{eq:nlarge}) is met, as $s_1^2s_2 \asymp \left( \frac{k \log(k)}{\varepsilon^2} \right)^2 \left( \frac{\log(k)}{\varepsilon^2} \right) \gtrsim
		\frac{n^2\log(k)}{\varepsilon^4}
		$. This holds by the definition of the regime.
		
		\item The $s_1$ constraint from (\ref{eq:nsmall}) requires $s_1 \gtrsim \frac{\sqrt{n\log(k)}+\log(k)}{\varepsilon^2}$, which is satisfied provided that $s_1 \gtrsim \frac{\log(k)}{\varepsilon^2}$ and $s_1 \gtrsim \frac{n}{\varepsilon^2}$ (since $\sqrt{n \log k}$ is the geometric mean of $n$ and $\log k$).  The former inequality is trivial, and the latter follows from the definition of the regime.
	\end{itemize}
	
	Combining these results gives the three distinct complexity regimes for the algorithm:
	\begin{itemize}
		\item For $ n \gtrsim \frac{k\log(k)}{\varepsilon^4}$, the complexity is $O\left(\left(\frac{n^2k\log(k)}{\varepsilon^4}\right)^{1/3}\right)$.
		\item For $k\log(k) \lesssim n \lesssim \frac{k\log(k)}{\varepsilon^4}$, the complexity is $O\left(\frac{\sqrt{nk\log(k)}}{\varepsilon^2}\right)$.
		\item For $n \lesssim k\log(k)$, the complexity is $O\left(\frac{k\log(k)}{\varepsilon^2}\right)$.
	\end{itemize}
\end{proof}

\subsection{Proof of Lemma~\ref{lem:esw_tester_correctness} ({\eswt} correctness)}\label{sec:esw_analysis}

In this section, we detail the {\eswt} procedure (Algorithm \ref{alg:empiricalsubsettest}) and prove its correctness guarantees stated in Lemma~\ref{lem:esw_tester_correctness}.
\begin{algorithm2e}[h!]
	\caption{$\eswt$}
	\label{alg:empiricalsubsettest}
	\DontPrintSemicolon
	\KwIn{$\varepsilon \in (0,1]$, sample access to distribution $P$, distributions $\{D_i\}_{i \in [k]}$, such that for $\chi \subseteq [k]$ we have $\{D_{i}\}_{i \in \chi} = P$, and $\{D_{i}\}_{i \in [k] \setminus \chi} = U$ with $\tv(P,U)\geq \varepsilon$.}
	\KwOut{A set $\hat{\chi} \subseteq [k]$ that, with probability at least $8/9$, satisfies $\hat{\chi} = \chi$.}
	$s_1 \gets \min\left(n,\frac{\sqrt{nk\log(k)}}{\varepsilon^2}\right)$\;
	$s_2 \gets c\frac{n\log(k)}{s_1\varepsilon^4}$ \tcp*{$c$ is a universal constant}
	$\mathcal{S} \sim P^{\otimes s_1}$\;
	$S \gets \mathrm{set}(\mathcal{S})$\;
	$\tau \gets \frac{s_1 \varepsilon^2}{64n}$\;
	$\hat{\chi} \gets \emptyset$\;
	\For{$j \in \{1,\ldots,k\}$}{
		$\mathcal{S}_j \sim D_j^{s_2}$\;
		$Z_j \gets \frac{1}{s_2} \sum_{i \in \mathcal{S}_j} \mathbbm{1}\{i \in S\}$\;
		\If{$Z_j \geq 1 - (1 - \frac{1}{n})^{s_1} + 2\tau$}{
			$\hat{\chi} \gets \hat{\chi} \cup j$\;
		}
	}
	\Return{$\hat{\chi}$}\;
\end{algorithm2e}

We restate the result as follows, and then proceed with the proof. 
\eswtcorrectness*
Let $\mathcal{S} = \{X_1, \dots, X_{s_1}\}$ be a multiset of $s_1$ samples drawn i.i.d. from a distribution $P$, a process we denote by $\mathcal{S} \sim P^{\otimes s_1}$. Let $S = \mathrm{set}(\mathcal{S})$ be the set of unique elements in $\mathcal{S}$ (i.e., duplicates are removed). 
The notation $\Pr_{\mathcal{S} \sim P^{\otimes s_1}}[\cdot]$ denotes the probability of an event where the probability measure is induced by the sampling of $\mathcal{S}$ from $P$.  Such events may be expressed in terms of $S = \mathrm{set}(\mathcal{S})$, in which case the contributions from all such $\mathcal{S}$ yielding that set are summed. 
Given $\mathcal{S} \sim P^{\otimes s_1}$, a result from~\citep{ACLS+22} establishes that the mass of $S$ is noticeably heavier in $P$ than in $U$.

We rely on a variation of this result from Lemma 2.18 of \citep{C22}. Our restatement omits the constants, and uses a tighter probability upper bound that only amounts to slight changes in those constants.

\begin{lemma}\label{lem:eswt} 
 {\em (Variation of Lemma 2.18 in \citep{C22})}
 For $P$ such that $\tv(P,U) \geq \varepsilon$ with $\varepsilon \gtrsim 1/n^{1/4}$, and for sets $S,S'$ constructed from $s_1$ samples, if $ \frac{\sqrt{n}}{\varepsilon^2} \lesssim s_1 \lesssim n$ (with suitable constants), then
 \begin{align*}
& \Pr_{\mathcal{S} \sim U^{\otimes s_1}}\left[U(S) \geq \mathbb{E}_{\mathcal{S}' \sim U^{\otimes s_1}}[U(S')] + \tau\right] \leq \frac{1}{36}\\
& \Pr_{\mathcal{S} \sim P^{\otimes s_1}}\left[P(S) \leq \mathbb{E}_{{\mathcal{S}' \sim U^{\otimes s_1}}}[U(S')] + 3\tau\right] \leq \frac{1}{36}
 \end{align*}
 for some $\tau$ satisfying $\tau \lesssim \frac{s_1\varepsilon^2}{n}$, where $U(S) = \sum_{x \in S} U(x)$, $P(S) = \sum_{x \in S} P(x)$, $S = \mathrm{set}(\mathcal{S})$, and $S' = \mathrm{set}(\mathcal{S}')$.
\end{lemma}

The following lemma can be summarized as ``non-uniform distributions tend to yield fewer distinct elements (more collisions) than the uniform distribution''.
\begin{restatable}{lemma}{majorise}\label{lem:majorise}
 For any $t \in \mathbb{R}$ and distribution $P$,
 \begin{align*}
\Pr_{\mathcal{S} \sim P^{\otimes s_1}}\left[U(S) \geq t\right] \leq
\Pr_{\mathcal{S} \sim U^{\otimes s_1}}\left[U(S) \geq t\right].
 \end{align*}
\end{restatable}
\begin{proof}
 The proof uses only elementary steps, and is deferred to Appendix~\ref{sec:majorise}.
\end{proof}

\begin{corollary}\label{cor:combo}
Under the setup of Lemma~\ref{lem:eswt}, we have
 \begin{align*}
\Pr_{\mathcal{S} \sim P^{\otimes s_1}}\left[U(S) \geq \mathbb{E}_{\mathcal{S}' \sim U^{\otimes s_1}}[U(S')] + \tau\right] \leq \frac{1}{36}.
 \end{align*}
\end{corollary}
\begin{proof}
 This follows from Lemmas~\ref{lem:eswt} and \ref{lem:majorise}.
\end{proof}
For a fixed set $\widetilde{S}$, the probability of a sample drawn from $D_j$ being in $\widetilde{S}$ is $D_j(\widetilde{S})$. We let $Z_j = \frac{1}{s_2} \sum_{t=1}^{s_2} \mathbbm{1}\{X_{t} \in \widetilde{S}\}$ denote an estimate of $D_j(\widetilde{S})$ based on $s_2$ samples from $D_j$. We observe that:
\begin{align} \label{eq:expcases}
 \mathbb{E}[Z_j] =
 \begin{cases}
U(\widetilde{S}) & \text{if } D_j = U \\
P(\widetilde{S})& \text{if } D_j = P.
 \end{cases}
\end{align} 
We now define two events for ${S}$ generated from $P$: $G_1$ is the event that $P({S}) \geq \mathbb{E}_{\mathcal{S}' \sim U}[U(S')] + 3\tau$, and $G_2$ the event that $U({S}) \leq \mathbb{E}_{\mathcal{S}' \sim U}[U(S')] + \tau$. We then have from Lemma~\ref{lem:eswt} that $\Pr[G_1]\geq 35/36$, and from Corollary~\ref{cor:combo} that $\Pr[G_2]\geq 35/36$.  By the union bound, we obtain $\Pr[G_1 \cap G_2] \geq 17/18$. We will henceforth assume $G_1 \cap G_2$ and suitably account for the total error probability later. 

We compute the expected mass (under $U$) of a set of $s_1$ samples from $U$ as $\mathbb{E}_{\mathcal{S}' \sim U^{\otimes s_1}}[U(S')] =  1 - (1 - \frac{1}{n})^{s_1}$ and we denote this as $\gamma$; we note for later use that $\gamma \le \frac{s_1}{n}$. Substituting the preceding findings into (\ref{eq:expcases}) (using $S$ for the generic set $\widetilde{S}$), we find that we have to distinguish between two cases for $Z_j$, for $\tau \lesssim \frac{s_1\varepsilon^2}{n}$:
\begin{align*}
    \mathbb{E}[Z_j] \leq \gamma + \tau  \quad \text{ and } \quad 
   \mathbb{E}[Z_j] \geq \gamma+ 3\tau  
\end{align*}
Since we have $k$ distributions, we seek to do this with error probability at most $\frac{1}{18k}$ per distribution. This is the same as determining with probability $1-\frac{1}{18k}$ whether, for a coin with unknown bias $\alpha$, we have $\alpha \leq \beta$ vs. $\alpha \geq \beta(1+\eta)$, where $\beta = \gamma + \tau$ and $\eta = \frac{2\tau}{\beta}$. By the multiplicative Chernoff bound, this requires $\Theta(\frac{\log(1/\delta)}{\beta\eta^2})$ samples for any error probability $\delta$. Since $\delta = \frac{1}{18k}$, and $\beta = \gamma + \tau \lesssim \frac{s_1}{n}\cdot\big(1+{\varepsilon^2} \big)\lesssim \frac{s_1}{n}$, and $\eta^2 = \frac{4\tau^2}{\beta^2}$, we find that the per-index sample complexity is $s_2  \asymp \frac{\log(18k)}{\beta\eta^2} \asymp  \frac{\log(k)n}{s_1\varepsilon^4}$. Let $E_3$ be the event that at least one of the tests fail. Then, by the union bound, $\Pr[E_3] \leq \frac{1}{18}$. The total failure probability of the tester is bounded by the failure of the ``good set" generation and the failure of the Chernoff bounds:
$ \Pr[\overline{G_1 \cap G_2}] + \Pr[E_3] \leq \frac{1}{18} + \frac{1}{18} = \frac{1}{9}$. Thus, the success probability is at least $\frac{8}{9}$.

The total sample complexity is $s_1 + k \cdot s_2$. We analyse the two cases from the lemma statement, which are determined by the definition of $s_1$:
\begin{itemize}
    \item \textbf{Case 1:} $n \gtrsim \frac{k\log(k)}{\varepsilon^4}$. In this case $s_1 = \frac{\sqrt{nk\log(k)}}{\varepsilon^2}$, and the complexity is
    $$ s_1 + k \cdot s_2 \asymp \frac{\sqrt{nk\log(k)}}{\varepsilon^2} + k \cdot \frac{n\log(k)}{s_1\varepsilon^4} \asymp \frac{\sqrt{nk\log(k)}}{\varepsilon^2} + \frac{kn\log(k)}{\frac{\sqrt{nk\log(k)}}{\varepsilon^2}\varepsilon^4} = O\left(\frac{\sqrt{nk\log(k)}}{\varepsilon^2}\right). $$
    \item \textbf{Case 2:} $n \lesssim \frac{k\log(k)}{\varepsilon^4}$. In this case $s_1 = n$, and the complexity is
    $$ s_1 + k \cdot s_2 = n + k \cdot \frac{n\log(k)}{n\varepsilon^4} = n + \frac{k\log(k)}{\varepsilon^4} = O\left(\frac{k\log(k)}{\varepsilon^4}\right), $$
    where the last step holds by the case condition.
\end{itemize}
Thus, we have established Lemma \ref{lem:esw_tester_correctness}.

\subsection{Proof of Theorem~\ref{thm:stage2_complexity_one_unknown} (Sample Complexity of Stage 2, One Known One Unknown)}\label{sec:stage2oneunknown}
    \oneknown*
    \begin{proof}
    Recall that for Stage 2, or the `Find All' phase, we had two possible algorithms, $\eswt$ and $\mlfht$. We choose the more efficient algorithm of the two, and that is decided based only on the domain size $n$. Hence, the total sample complexity is the minimum of the sample complexity of the two. 
    
    The total sample complexity for the $\eswt$ is given by: 
        \begin{align*}
            T_{\mathtt{ESW}} =
        \begin{cases}
            O\left(\frac{\sqrt{nk\log(k)}}{\varepsilon^2}\right) & \text{if } n \gtrsim \frac{k\log(k)}{\varepsilon^4} \\
            O\left(\frac{k\log(k)}{\varepsilon^4}\right) & \text{if } n \lesssim \frac{k\log(k)}{\varepsilon^4}
        \end{cases} \qquad \qquad \text{(From Lemma~\ref{lem:esw_tester_correctness})}
        \end{align*}
        The complexity for $\mlfht$,  is:
        $$ T_{\mathtt{Mul}} = 
        \begin{cases}
            O\left(\left(\frac{n^2k\log(k)}{\varepsilon^4}\right)^{1/3}\right) & \text{if } n \gtrsim \frac{k\log(k)}{\varepsilon^4} \\
            O\left(\frac{\sqrt{nk\log(k)}}{\varepsilon^2}\right) & \text{if } k\log(k) \lesssim n \lesssim \frac{k\log(k)}{\varepsilon^4} \\
            O\left(\frac{k\log(k)}{\varepsilon^2}\right) & \text{if } n \lesssim k\log(k)
        \end{cases} \qquad \qquad \text{(From Theorem~\ref{thm:multilfht})}
        $$
        Recall that Lemma \ref{lem:esw_tester_correctness} (on $\eswt$) requires $n \gtrsim \frac{1}{\varepsilon^4}$; we will ensure this condition below. The overall sample complexity of Stage 2 is $T_{2} = \min(T_{\mathtt{ESW}},T_{\mathtt{Mul}})$, which we study through a case analysis.
        
        \begin{itemize}
            \item \textbf{Case 1: Large Domain ($n \gtrsim \frac{k\log(k)}{\varepsilon^4}$)}.
            We compare $T_{\mathtt{ESW}} = O\left(\frac{\sqrt{nk\log(k)}}{\varepsilon^2}\right)$ with $T_{\mathtt{Mul}} = O\left(\left(\frac{n^2 k \log(k)}{\varepsilon^4}\right)^{1/3}\right)$. To simplify, we compare their 6th powers.
            \begin{align*}
                (T_{\mathtt{ESW}})^6 &\propto \left(\frac{\sqrt{nk\log(k)}}{\varepsilon^2}\right)^6 = \frac{k^3 n^3 (\log k)^3}{\varepsilon^{12}} \\
                (T_{\mathtt{Mul}})^6 &\propto \left(\left(\frac{n^2 k \log(k)}{\varepsilon^4}\right)^{1/3}\right)^6 = \frac{n^4 k^2 (\log k)^2}{\varepsilon^8}
            \end{align*}
            Dividing both sides by the common term $\frac{n^3 k^2 (\log k)^2}{\varepsilon^8}$ shows that this comparison is equivalent to comparing $\frac{k \log k}{\varepsilon^4}$ with $n$. Since $n \gtrsim \frac{k \log k}{\varepsilon^4}$ in this regime, it follows that $T_{\mathtt{ESW}} \lesssim T_{\mathtt{Mul}}$, i.e., $\eswt$ is the better choice.
            
            \item \textbf{Case 2: Intermediate and Small Domains ($n \lesssim \frac{k\log(k)}{\varepsilon^4}$)}.
            We compare the relevant complexities for $T_{\mathtt{Mul}}$ and $T_{\mathtt{ESW}}$ in this range by splitting it based on the boundaries for $T_{\mathtt{Mul}}$:
            \begin{itemize}
                \item If $k\log(k) \lesssim n \lesssim \frac{k\log(k)}{\varepsilon^4}$: We compare $T_{\mathtt{ESW}} = O\left( \frac{k\log(k)}{\varepsilon^4}\right)$ with $T_{\mathtt{Mul}} = O\left(\frac{\sqrt{nk\log(k)}}{\varepsilon^2}\right)$. Plugging the maximum value of $n$ in this regime, $n \asymp \frac{k\log(k)}{\varepsilon^4}$, we see that $\mlfht$ is more efficient over the entire regime.
                \item If $n \lesssim k\log(k)$: We compare $T_{\mathtt{ESW}} = O\left(\frac{k\log(k)}{\varepsilon^4}\right)$ with $T_{\mathtt{Mul}} = O\left(\frac{k\log(k)}{\varepsilon^2}\right)$. The $1/\varepsilon^4$ term in $T_{\mathtt{ESW}}$ ensures it is larger than $T_{\mathtt{Mul}}$.
            \end{itemize}
            Hence in both sub-cases of Case 2, $\mlfht$ is the better choice.
        \end{itemize}
    
        By combining these results, we find that the complexity from Case 1  and the first part of Case 2  merge into a single regime for $n \gtrsim k\log(k)$. The second part of Case 2 covers $n \lesssim k\log(k)$. The complexity for Stage 2 is therefore:
        \begin{align*} 
        \begin{cases}
            O\left(\frac{\sqrt{nk\log(k)}}{\varepsilon^2}\right) & \text{if } n \gtrsim k\log(k) \\[2ex]
            O\left(\frac{k\log(k)}{\varepsilon^2}\right) & \text{if } n \lesssim k\log(k)
        \end{cases}
        \end{align*}
    \end{proof}
    
    \subsection{Sample Complexity of Clustering (One Known One Unknown)}\label{sec:derivation_oneknown}
    \begin{restatable}{theorem}{oneknownupper}\label{thm:one_known_upper}
     Our algorithm solves the 2-clustering problem with one known distribution with success probability at least $\frac{2}{3}$. The total sample complexity is given by:
     \begin{itemize}
        \item For $n \gtrsim k\log(k)$:
        $$ O\left( \frac{k\log(k)\sqrt{n}}{r\varepsilon^2} + \frac{\sqrt{nk\log(k)}}{\varepsilon^2} \right).$$
        
        \item For $n \lesssim k\log(k)$:
        $$O\left( \frac{k\log(k)\sqrt{n}}{r\varepsilon^2} + \frac{k\log(k)}{\varepsilon^2} \right).$$
     \end{itemize}
    \end{restatable}
    \begin{proof}
    The algorithm for this case proceeds in two stages, and the total sample complexity is the sum of their costs, $T = T_1 + T_2$. We have the component complexities as follows:
    \begin{align*}
        T_1 &= O\left( \frac{k\log(k)\sqrt{n}}{r\varepsilon^2} \right) \qquad\qquad {\text{(From Stage 1 in Section~\ref{sec:1known1unknown})}} \\
        T_2  &= 
        \begin{cases}
            O\left(\frac{\sqrt{nk\log(k)}}{\varepsilon^2}\right) & \text{if } n \gtrsim k\log(k) \\[2ex]
            O\left(\frac{k\log(k)}{\varepsilon^2}\right) & \text{if } n \lesssim k\log(k)
        \end{cases} \qquad\qquad\text{(From Theorem~\ref{thm:stage2_complexity_one_unknown})}
    \end{align*}
    The bounds follow by summing $T_1$ and $T_2$ for each of the two corresponding regimes for $n$.
    \end{proof}
    
\subsection{Finding One Distribution in Many}\label{sec:unknownunknownupperbound}

Our procedure for finding one unknown distribution is given in Algorithm \ref{alg:uneqequitest}.  Since we will apply this procedure on a suitably chosen subset of our distributions rather than the full set of distributions, we use generic notation $m$ for the number of distributions here.

\begin{algorithm2e}[h!]
	\caption{Find-One-Unknown}
	\label{alg:uneqequitest}
	\DontPrintSemicolon
	\LinesNumbered
	\KwIn{$\varepsilon \in (0,1)$, sample access to distributions $\{D_i\}_{i=1}^{m}$.}
	\KwOut{A pair of indices $\mathcal{E} \subseteq [m]$ representing one exemplar from each cluster, or a report that all distributions are identical.}
		 $s_1 \gets \min(n,(\frac{nm}{\varepsilon^{2}})^{2/3})$\;
		 $\delta' \gets \frac{1}{30m}$\;
		 $b \gets \sqrt{\frac{30}{s_1}}$\;
		 $s_2 \gets Cn s_1^{-1/2} \frac{\log(1/{\delta'})}{\varepsilon^2}$ \tcp*{$C$ is a universal constant}
		 $s \sim \mathrm{Poi}(s_1)$ \tcp*{$\Pr[\mathrm{Poi}(\mu)=j] = \frac{e^{-\mu}\mu^j}{j!}$}
		\If{$s > 2s_1$}{\Return $\bot$}
		Define a multiset $\mathcal{S}$ by taking $s$ samples from $D_1$\;
		\For{$i \in [m] \setminus \{1\}$}{
			Run the algorithm from Lemma~\ref{lem:L2test} for $D_{1,\mathcal{S}}, D_{i,\mathcal{S}}$ with $s_2$ samples, $\varepsilon,\delta'$, and $b$\;
			\If{the test returns $\tv(D_{1,\mathcal{S}}, D_{i,\mathcal{S}}) > \varepsilon $}{ \Return $\{1,i\}$}
		}
		\Return \text{All Identical}
\end{algorithm2e}

In this section, we will prove the following theorem, and the proof closely follows the proof of Proposition~2.11 in \citep{DK16}.
\begin{theorem}\label{thm:main_equiv_split_test}
	Find-One-Unknown (Alg.~\ref{alg:uneqequitest}) takes in $m \geq 2$ distributions $\{D_i\}_{i \in [m]}$ over $[n]$ (for $n\geq 27$), and parameters $\varepsilon,\delta \in (0,1)$, such that for an unknown set $\chi \subseteq [m]$, and unknown distributions $P$ and $Q$, with $\tv(P,Q) \geq \varepsilon$, $D_{i} = P$ for $i \in \chi$, and $D_{i}= Q$ for $ i \in [m]\setminus\chi$.
	With probability at least $2/3$, if $\chi = [m]$, the algorithm returns 'All  Identical', and if $\chi \neq [m]$, it outputs an index $j$ such that $D_j \not = D_1$. The sample complexity is 
	$$O\left(\max\left(\left(\frac{nm}{\varepsilon^2}\right)^{2/3}, \frac{m\sqrt{n}}{\varepsilon^2}\right)\log(m)\right)$$
\end{theorem}

We use the following result from \cite{DK16} where it appears as Lemma 2.3. The original lemma stated a success probability of $\frac{2}{3}$, but we state it in terms of $1-\delta$ where $\delta \in (0,1)$. This is possible via a majority vote, which can boost the success probability from $\frac{2}{3}$ to $1-\delta$ at a multiplicative cost of $O(\log(\frac{1}{\delta}))$ in the sample complexity.
\begin{lemma}\label{lem:L2test}
	Let $P$ and $Q$ be two unknown distributions on $[n]$. There exists an algorithm that on inputs $\varepsilon,\delta \in (0,1)$, and $b \ge \min \{||P||_2, ||Q||_2\}$, draws $O(bn\log(\frac{1}{\delta})/\varepsilon^2)$ samples from each of $P$ and $Q$ and, with probability at least $1-\delta$, distinguishes between $\tv(P,Q) = 0$ and $\tv(P,Q) \geq \varepsilon$.
\end{lemma}

\begin{proof}[Proof of Theorem~\ref{thm:main_equiv_split_test}]
	The algorithm draws $s$ samples from $D_1$ to create the multiset $\mathcal{S}$, where $s$ is a random variable drawn from a Poisson distribution with mean $s_1$. If $s> 2s_1$, the algorithm terminates immediately and returns $\bot$. Let $E_1$ denote the event that $s> 2s_1$. By the  Chernoff bound and $s_1 \ge n^{2/3}$, we have $\Pr[E_1] = \Pr[s > 2s_1] \le (\frac{e}{4})^{s_1} \leq (\frac{e}{4})^{n^{2/3}}  \leq \frac{1}{30}$ (for $n \ge 27$).
	
	As laid out in \citep{DK16}, the multiset $\mathcal{S}$ is used to transform all of the $\{D_i\}_{i \in [m]}$ into new ``flattened'' distributions $\{D_{i,\mathcal{S}}\}_{i \in [m]}$ satisfying the following properties:
	\begin{itemize}
		\item These distributions $D_{i,\mathcal{S}}$ are defined on an expanded domain of size $n + |\mathcal{S}|$;
		\item The pairwise TV distances of these distributions are identical to those of $\{D_i\}_{i \in [m]}$ (see Fact~2.5 of \citep{DK16});
		\item These distributions have low $\ell_2$ norm on average; specifically, $\mathbb{E}[\|D_{i,\mathcal{S}}\|_2^2] \le \frac{1}{s_1}$ (see Lemma~2.6 of \citep{DK16}). 
	\end{itemize}
	Let $E_2$ denote the event that $\|D_{1,\mathcal{S}}\|_2 > \sqrt{\frac{30}{s_1}}$. By Markov's inequality, $$\Pr[E_2] = \Pr\left[\|D_{1,\mathcal{S}}\|_2 > \sqrt{\frac{30}{s_1}}\right] = \Pr\left[\|D_{1,\mathcal{S}}\|_2^2 > \frac{30}{s_1}\right] \le \frac{\mathbb{E}[\|D_{1,\mathcal{S}}\|_2^2]}{30/s_1} \le  \frac{1}{30}.$$
	
	We perform our analysis conditioned on $\overline{E}_2$ (the complement of $E_2$), i.e., $\|D_{1,\mathcal{S}}\|_2 \le \sqrt{\frac{30}{s_1}}$. The algorithm sets $b = \sqrt{\frac{30}{s_1}}$. Thus, on this event, the prerequisite condition $b \ge \min (\|D_{1,\mathcal{S}}\|_2, \|D_{i,\mathcal{S}}\|_2)$ for Lemma~\ref{lem:L2test} is satisfied, since $\min (\|D_{1,\mathcal{S}}\|_2, \|D_{i,\mathcal{S}}\|_2) \le \|D_{1,\mathcal{S}}\|_2 \le b$. The tester then distinguishes between $\tv(D_{1,\mathcal{S}},D_{i,\mathcal{S}}) = 0$ and $\tv(D_{1,\mathcal{S}}, D_{i,\mathcal{S}}) \ge \varepsilon$, which is equivalent to distinguishing between $\tv(D_{1},D_{i}) = 0$ and $\tv(D_{1}, D_{i}) \ge \varepsilon$. The tester draws $s_2 = Cb n \log(1/\delta')/\varepsilon^2 \asymp s_1^{-1/2}n\log(\frac{1}{\delta'})/\varepsilon^2$ samples, and errs with probability at most $\delta'$. Let $E_3$ be the event that at least one of the $m-1$ tests fail. Since $\delta' = \frac{1}{30m}$, a union bound over the $m-1$ tests limits the total error to be at most $\Pr[E_3] \le (m-1)\delta' \le \frac{m-1}{30m} \le \frac{1}{30}$. The total probability of error is at most $\Pr[E_1] + \Pr[E_2] + \Pr[E_3] \leq \frac{1}{10}$, by the union bound.
	
	We condition on the success event $\overline{E} = \overline{E}_1 \cap \overline{E}_2 \cap \overline{E}_3$, which happens with probability at least $1-\frac{1}{10} = \frac{9}{10}$. If $\chi = [m]$, then since the test from Lemma~\ref{lem:L2test} run correctly, the loop completes and the algorithm returns ``All Identical''. If $\chi \not= [m]$, then there exists at least one $i \in [m]\setminus \{1\}$ such that $\tv(D_1, D_i) \ge \varepsilon$. Since the test does not fail, it will correctly return that index $i$.
	
	The total sample complexity, $T = O(s_1 + m \cdot s_2)$, is determined by the sketching cost ($s_1$) and the total comparison cost ($m \cdot s_2$). The algorithm chooses $s_1 = \min(n, (\frac{nm}{\varepsilon^2})^{2/3})$ to balance these costs. In more detail, the per-comparison cost is $s_2 \asymp n s_1^{-1/2} \cdot \frac{\log(m)}{\varepsilon^2}$, and this creates two regimes:
	\begin{itemize}
		\item When $m < \varepsilon^2\sqrt{n}$, the algorithm balances the costs by setting $s_1 = (\frac{nm}{\varepsilon^2})^{2/3}$. This results in a total complexity of $O\left(\left(\frac{nm}{\varepsilon^2}\right)^{2/3}\log(m)\right)$.
		\item When $m \ge \varepsilon^2\sqrt{n}$, we cap the sketching cost at $s_1=n$ to prevent the expanded domain size $n + |\mathcal{S}|$ (with $|\mathcal{S}| = O(s_1)$) from being significantly higher than the original size $n$.\footnote{In principle we could allow $n + |\mathcal{S}| \gg n$, but it can be verified that doing so is not beneficial here.}   
		The comparison cost dominates, leading to a total complexity of $O\left(\frac{m\sqrt{n}}{\varepsilon^2}\log(m)\right)$.
	\end{itemize}
	The overall complexity is the maximum of the resulting terms from these two regimes:
	$$O\left(\max\left(\left(\frac{nm}{\varepsilon^2}\right)^{2/3}, \frac{m\sqrt{n}}{\varepsilon^2}\right)\log(m)\right).$$
\end{proof}

\subsection{Sample Complexity of Clustering (Both Unknown)}\label{sec:derivationbothunknown}
    \begin{restatable}{theorem}{bothunknownupper}\label{thm:both_unknown_upper}
     Our algorithm solves the 2-clustering problem with two unknown distributions with success probability at least $\frac{2}{3}$. The total sample complexity is given by:
     \begin{itemize}
        \item For $n \gtrsim \frac{k\log(k)}{\varepsilon^4}$:
        $$ O\left(\max\left(\left(\frac{nk}{r\varepsilon^2}\right)^{2/3}, \frac{k\sqrt{n}}{r\varepsilon^2}\right)\log(k) + \left(\frac{n^2k\log(k)}{\varepsilon^4}\right)^{1/3}\right).$$
        
        \item For $k\log(k) \lesssim n \lesssim \frac{k\log(k)}{\varepsilon^4}$:
        $$O\left(\max\left(\left(\frac{nk}{r\varepsilon^2}\right)^{2/3}, \frac{k\sqrt{n}}{r\varepsilon^2}\right)\log(k) + \frac{\sqrt{nk\log(k)}}{\varepsilon^2}\right).$$
        
        \item For $n \lesssim k\log(k)$:
        $$O\left(\max\left(\left(\frac{nk}{r\varepsilon^2}\right)^{2/3}, \frac{k\sqrt{n}}{r\varepsilon^2}\right)\log(k) + \frac{k\log(k)}{\varepsilon^2}\right).$$
     \end{itemize}
    \end{restatable}
    \begin{proof}
    This algorithm proceeds in two stages, with the total complexity given by the sum of their costs, $T = T_1 + T_2$. For the first stage we call Find-One-Unknown (Algorithm~\ref{alg:uneqequitest}) with $3k/r$ subsampled arms, i.e.,  we set $m \gets 3k/r$.  Thus we get following component complexities:
    \begin{align*}
        T_1 &= O\left(\max\left(\left(\frac{nk}{r\varepsilon^2}\right)^{2/3}, \frac{k\sqrt{n}}{r\varepsilon^2}\right)\log(k/r)\right) \qquad \qquad \text{(From Theorem~\ref{thm:main_equiv_split_test})}\\
        T_2 &= 
        \begin{cases}
            O\left(\left(\frac{n^2k\log(k)}{\varepsilon^4}\right)^{1/3}\right) & \text{if } n \gtrsim \frac{k\log(k)}{\varepsilon^4} \\[2ex]
            O\left(\frac{\sqrt{nk\log(k)}}{\varepsilon^2}\right) & \text{if } k\log(k) \lesssim n \lesssim \frac{k\log(k)}{\varepsilon^4} \\[2ex]
            O\left(\frac{k\log(k)}{\varepsilon^2}\right) & \text{if } n \lesssim k\log(k)
        \end{cases}\qquad \qquad \text{(From Theorem~\ref{thm:multilfht})}
    \end{align*}
    The theorem's bounds follow by summing $T_1$ and $T_2$ for each of the three corresponding regimes for $n$.
    \end{proof}

\input{extension}

\section{Derivation of Lower Bounds}\label{sec:appendix_lower_bounds}
    
    \subsection{Proof of Theorem~\ref{thm:indexfindinglb} (Unequal Sample Testing Based Lower Bounds)}\label{sec:testinglb}
    \indexfindinglb*
    
    \begin{proof}
    	The proof proceeds via a reduction. We first establish in the following lemma that any algorithm \(A\) capable of identifying an index from the unknown set \(\mathcal{I}\) can also be used to solve the \(\UST\) problem defined in Section~\ref{sec:generalised}. Recall that \(\UST(P,Q,s_1,s_2)\) is the task of distinguishing, with probability at least \(2/3\), between the hypotheses \(H_0: P = Q\) and \(H_1: \operatorname{TV}(P,Q) \ge \varepsilon\), given \(s_1\) samples from \(P\) and \(s_2\) samples from \(Q\). The lower bounds for \(\UST(P,Q,s_1,s_2)\) (Lemma~\ref{lem:unequal}) are proved by constructing two explicit distributions: \(\Lambda_{\mathrm{alt}}\), supported on pairs \((P,Q)\) that are \(\varepsilon\)-far, and \(\Lambda_{0}\), supported on pairs \((P,Q)\) with \(P = Q\), such that no algorithm can  distinguish (with at least $0.9$ probability) whether \((P,Q) \sim \Lambda_{\mathrm{alt}}\) or \((P,Q) \sim \Lambda_{0}\) unless the number of samples $s_1$ and $s_2$ satisfies as mentioned in Lemma \ref{lem:unequal}.
    
    For example, the first part of Lemma~\ref{lem:unequal}, i.e, when $Q=U$, is proved in \cite{P08} by taking \(\Lambda_0 = \{(U,U)\}\), and letting \(\Lambda_{\mathrm{alt}} = (U,Q)\) where $Q$ is the  distribution chosen uniformly at random from the family of non-uniform distributions \(\{Q_S : S \subseteq [n],\, |S| = n/2\}\), where each \(Q_S\) is defined as
    \[
    Q_S(i) =
    \begin{cases}
    \dfrac{1+2\varepsilon}{n}, & i \in S,\\[6pt]
    \dfrac{1-2\varepsilon}{n}, & i \notin S,
    \end{cases}
    \qquad i \in [n].
    \]
    
    Similarly, the proof of the second part of Lemma~\ref{lem:unequal} relies on constructing two explicit distributions, $\Lambda_{\mathrm{alt}}$ and $\Lambda_0$. For the purposes of proving Theorem~\ref{thm:indexfindinglb}, we only require the \emph{existence} of these distributions as a black box; their explicit descriptions are not needed. Therefore, we do not define the concrete constructions for the second case here, it can be found in \cite{BV15}. The proof of Theorem \ref{thm:indexfindinglb} relies on the following lemma.
    
    	\begin{restatable}{lemma}{genredindex}\label{lem:general-reduction-to-testing-index}
    		Suppose that there exists a randomized algorithm $A$ as stated in Theorem~\ref{thm:indexfindinglb}. 
            Let $A$ draws a total of $T$ samples. Then there exists an algorithm $A'$ that solves the $\UST$ problem (distinguishing $(P_1,P_2) \sim \Lambda_0$ from $(P_1,P_2) \sim \Lambda_\mathrm{alt}$) with probability $\ge 2/3$, using at most $s_1 \le T$ samples from $P_1$ and $s_2 \le 60rT/k$ samples from $P_2$.
    	\end{restatable}
    	
    	Before proving the above lemma, we show how this lemma implies Theorem \ref{thm:indexfindinglb}. We  apply the known lower bounds for the $\UST(P_1,P_2,s_1,s_2)$ problem to the sample complexities $s_1 \le T$ and $s_2 \le 60rT/k$ of algorithm $A'$.
    	\paragraph{Part (a): One Known.} 
        The relevant lower bound from Lemma~\ref{lem:unequal} for this testing problem is $s_{2} \gtrsim \frac{\sqrt{n}}{\varepsilon^2}$. Substituting into $s_2$: $s_2 \gtrsim \frac{\sqrt{n}}{\varepsilon^2} \implies  \frac{60rT}{k} \gtrsim \frac{\sqrt{n}}{\varepsilon^2} \implies T \gtrsim \frac{k\sqrt{n}}{r\varepsilon^2}$. The required number of samples is $T \in \Omega(\frac{k\sqrt{n}}{r\varepsilon^2})$.
    	
    	\paragraph{Part (b): Both Unknown.}
        The bounds from Lemma~\ref{lem:unequal} are split into two regimes.
    	\begin{itemize}
    		\item \textbf{Regime 1: $n \lesssim 1/\varepsilon^4$.} The bounds are identical to Part (a), yielding a lower bound of $T \gtrsim \frac{k\sqrt{n}}{r\varepsilon^2}$.
    		
    		\item \textbf{Regime 2: $n \gtrsim 1/\varepsilon^4$.} The relevant bounds are $s_1s_{2}^2  \gtrsim (\frac{n}{\varepsilon^2})^2$.
    		Substituting $s_1$ and $s_2$:
    		\begin{itemize}
    			\item $s_1 s_2^2  \gtrsim (\frac{n}{\varepsilon^2})^2 \implies \left(\frac{60rT}{k}\right)^2 \cdot T \gtrsim \frac{n^2}{\varepsilon^4} \implies T^3 \gtrsim \frac{n^2 k^2}{r^2 \varepsilon^4} \implies T \gtrsim \left(\frac{nk}{r\varepsilon^2}\right)^{2/3}$.
    		\end{itemize}
    	\end{itemize}
    	In the $n \gtrsim 1/\varepsilon^4$ regime, the $(\frac{nk}{r\varepsilon^2})^{2/3}$ bound dominates the $(\frac{k\sqrt{n}}{r\varepsilon^2})^{1/2}$ bound. Therefore, $T \in \Omega(\max \{ \frac{k \sqrt{n}}{r\varepsilon^{2}}, (\frac{nk}{r \varepsilon^{2}})^{2/3} \})$.
    \end{proof}
    
    \subsubsection{Proof of Lemma~\ref{lem:general-reduction-to-testing-index}}
        \begin{proof}
        To describe the algorithm $A'$, we will first construct a set of indices $J$ with some nice properties. This set $J$ will depend on the algorithm $A$ and the distribution $\Lambda_0$.
        
        We construct  a random input instance $\Delta_0$ as follows: we sample  $(P_1,P_2) \sim {\Lambda_0}$. Then we sample a uniformly random subset $\mathcal{I}_0 \sim [k]$ such that $|\mathcal{I}_0| = r$. Then we set $\{D_{i}\}_{i \in \mathcal{I}_0} = P_2$, and $\{D_{i}\}_{i \in [k]\setminus \mathcal{I}_0} = P_1$.
    	
            Let $N_a^0$ be the random variable for the number of times distribution $a$ is sampled from by $A$ under $\Delta_0$, and let $p_a^0$ be the probability that $A$ outputs index $a$ under $\Delta_0$. The expectations and probabilities are over the drawing of $(P_1,P_2)$, $A$'s internal randomness and the samples drawn.
    		We have $\sum_{a=1}^k \mathbb{E}[N_a^0] = T$ and $\sum_{a=1}^k p_a^0 = 1$.
    		
    		Define three sets of indices:
    		\begin{itemize}
    			\item $J_1 = \{a \mid \mathbb{E}[N_a^0] \le 3T/k \}$
    			\item $J_2 = \{a \mid p_a^0 \le 3/k \}$
    			\item $J = J_1 \cap J_2$
    		\end{itemize}
    		Since $\sum_{a=1}^k \mathbb{E}[N_a^0] = T$, at most $k/3$ indices can satisfy $\mathbb{E}[N_a^0] > 3T/k$. Thus, $|J_1| \ge 2k/3$. Similarly, since $\sum_{a=1}^k p_a^0 = 1$, at most $k/3$ indices can satisfy $p_a^0 > 3/k$, and hence $|J_2| \ge 2k/3$. Then $|J| = |J_1| + |J_2| - |J_1 \cup J_2| \ge k/3$.
    		Let $M_0 = 60rT/k$. For any set $I' \subseteq J$ of size $|I'| \leq r$, we have
    		\begin{align}
    			\Pr\left[\sum_{j \in I'} N_j^0 > M_0\right] \le \frac{\mathbb{E}\left[\sum_{j \in I'} N_j^0\right]}{M_0} = \frac{\sum_{j \in I'} \mathbb{E}\left[N_j^0\right]}{M_0}  \leq  \frac{|I'|\cdot 3T/k}{M_0} \leq \frac{3rT/k}{60rT/k} =  0.05 \label{eq:markovbound_abstract}
    		\end{align}
    		Now, we construct an algorithm $A'$ (using $A$) to solve the testing problem (distinguishing $(P_1,P_2) \sim \Lambda_0$ from $(P_1,P_2) \in \Lambda_\mathrm{alt}$). $A'$ works as follows:
    		\begin{enumerate}
    			\item $A'$ takes as input a pair of distributions $(P_1,P_2)$ (which is either drawn from $\Lambda_0$ or $\Lambda_\mathrm{alt}$).
    			\item $A'$ picks $r$ indices arbitrarily from the set $J$, call it $I' \subseteq J$. It sets up an instance for $A$: for $j \in I'$, $D_j = P_2$, and for $j \in [k]\setminus I'$, $D_j = P_1$.
    			\item $A'$ simulates $A$ on this instance. Let $N_{j}$ be the number of times $A$ draws a sample from $D_j$, and let $Y_A$ be the index output by $A$.
    			\item If, at any point in the simulation, $\sum_{j \in I'} N_{j} > M_0$, $A'$ aborts  and outputs $\Lambda_\mathrm{alt}$.
    			\item Otherwise ($\sum_{j \in I'} N_{j} \le M_0$) $A'$ outputs:
    			\begin{itemize}
    				\item $\Lambda_\mathrm{alt}$, if ($Y_A \in I'$).
    				\item $\Lambda_{0}$, if ($Y_A \not \in I'$).
    			\end{itemize}
    		\end{enumerate}
    		We now show that $A'$ correctly classifies $(P_1,P_2) \sim \Lambda_0$ or $\Lambda_\mathrm{alt}$ with at least $0.9$ probability.
    		\paragraph{Case 1: $(P_1,P_2) \sim \Lambda_{0}$.} This corresponds to the $\Delta_0$ scenario for $A$. $A'$ is in error if it outputs $\Lambda_\mathrm{alt}$. This occurs if ($\sum_{j \in I'} N_j^0 > M_0$) or ($\sum_{j \in I'} N_j^0 \le M_0$ and $Y_A \in I'$). Since these two events are disjoint,
    		\begin{align*}
    			\Pr[\text{error}|(P_1,P_2) \sim \Lambda_{0} ] 
    			&= \Pr\left[\sum_{j \in I'}N_j^0 > M_0\right] +\Pr\left[\left(Y_A \in I' \right)\wedge \left(\sum_{j \in I'}N_j^0 \le M_0\right)\right]\\
    			&\leq  \underbrace{0.05}_{\text{from eq.}(\ref{eq:markovbound_abstract})} +\Pr\left[Y_A \in I' \right] \\
    			&\leq  0.05 + \sum_{j \in I'}p_j^0 \le 0.05 +  \frac{3r}{k} \leq 0.05+\underbrace{0.05}_{r \leq k/60} = 0.1.
    		\end{align*}
    		
    		\paragraph{Case 2: $(P_1,P_2) \sim \Lambda_\mathrm{alt}$.}
    		Recall $I'$ defined in step 2 above, and that we set $D_j = P_2$ for $j \in I'$, and $D_j = P_1$ for $j \in [k] \setminus I'$. This is an instance $A$ is designed for. $A'$ errs if it outputs "$(P_1,P_2)\in \Lambda_0$". This happens if ($\sum_{j \in I'}N_{j} \leq M_0$ and $Y_A \not \in I'$), so, $\Pr[\text{error}\,|\,(P_1,P_2)\sim {\Lambda_\mathrm{alt}} ] \leq \Pr[Y_A \not \in I']$. By property (3) of $A$, $\Pr[Y_A  \in I'] \geq 0.9$, and thus $\Pr[\text{error}\,|\,(P_1,P_2)\sim {\Lambda_\mathrm{alt}} ] \leq \Pr[Y_A \not \in I'] \le 0.1$.
    		
    		Therefore, if $r \le k/60$, algorithm $A'$ solves the testing problem (distinguishing $\Lambda_0$ from $\Lambda_\mathrm{alt}$) with success probability at least $0.9$. It uses $s_1 = \sum_{j \in [k]\setminus I'} N_j \le T$ samples from $P_1$  and $s_2 = \sum_{j \in I'} N_j \le M_0 = 60rT/k$ samples from $P_2$.
    	\end{proof}

\subsection{Proof of Theorem~\ref{thm:combined-lb-lfht} (LFHT Based Lower Bounds)}\label{sec:proofoflfhtlb}
\lfhtlb*

The lower bound for \(\LFHT(P, Q, s_1, s_2)\) (Lemma~\ref{lem:lfht}) is proved by constructing an explicit distribution \(\Lambda_{\mathrm{alt}}\) supported on pairs \((P,Q)\) that are \(\varepsilon\)-far. It is shown that, for \((P,Q) \sim \Lambda_{\mathrm{alt}}\), no algorithm can distinguish, with probability at least \(0.9\), whether a given unknown distribution \(D \in \{P,Q\}\), is \(P\) or \(Q\) unless the number of samples $s_1$ and $s_2$ satisfies as mentioned in Lemma \ref{lem:lfht}.
 Again, as in Section~\ref{sec:testinglb}, we will make use of  $\Lambda_{\mathrm{alt}}$ in a ``black box'' manner, and we therefore do not describe their constructions here. Their detailed definitions can be found in~\cite{GHP23}.

\begin{lemma}\label{lem:general-reduction-to-lfht}
	Suppose there exists a randomized algorithm $A$ satisfying the conditions in Theorem~\ref{thm:combined-lb-lfht}. Let $A$ draw a total of $T$ samples. 
	Then there exists an algorithm $A'$ that solves the $\LFHT(P, Q, s_1, s_2)$ problem when $(P,Q) \sim \Lambda_{\mathrm{alt}}$, using $s_1 \le T$ samples from $P$ and $Q$ each, and $s_2 \le \frac{80T}{k}$ samples from $D$, i.e., $A'$ solves $\LFHT(P,Q,T,\frac{80T}{k})$ when $(P,Q) \sim \Lambda_{\mathrm{alt}}$.
\end{lemma}
Before proving the above lemma, we show that it implies Theorem \ref{thm:combined-lb-lfht}.
\paragraph*{Part (a): One Known}
We apply Lemma~\ref{lem:general-reduction-to-lfht} setting $Q = U$.
The premise of Theorem~\ref{thm:combined-lb-lfht} (Part 1) assumes an algorithm $A$ satisfying the conditions of Lemma~\ref{lem:general-reduction-to-lfht} for these distributions.
The lemma guarantees an algorithm $A'$ solving $\LFHT(P,U,s_1,s_2)$ with $s_1 \le T$ samples from $P$ and $s_2 \le \frac{80T}{k}$ samples from $D$.
From Lemma~\ref{lem:lfht}, such an algorithm must satisfy $s_1 \gtrsim \frac{\sqrt{n}}{\varepsilon^2}$, $s_1s_2 \gtrsim \frac{n}{\varepsilon^4}$, and $s_2 \gtrsim \frac{1}{\varepsilon^2}$.
Substituting $s_1 \le T$ and $s_2 \le \frac{80T}{k}$:
\begin{itemize}
	\item $T \gtrsim s_1 \gtrsim \frac{\sqrt{n}}{\varepsilon^2}$
	\item $T (\frac{80T}{k}) \gtrsim s_1s_2 \gtrsim \frac{n}{\varepsilon^4} \implies T \gtrsim \frac{\sqrt{nk}}{\varepsilon^2}$
	\item $\frac{80T}{k} \gtrsim s_2 \gtrsim \frac{1}{\varepsilon^2} \implies T \gtrsim \frac{k}{\varepsilon^2}$
\end{itemize}
Combining these, we get $T \gtrsim \frac{k+\sqrt{nk}}{\varepsilon^2}$.

\paragraph*{Part (b): Both Unknown}
We apply Lemma~\ref{lem:general-reduction-to-lfht} with unknown distributions $P$ and $Q$.
The premise of Theorem~\ref{thm:combined-lb-lfht} (Part 2) assumes an algorithm $A$ that identifies the $k-r$ $P$ distributions from the $r$ $Q$ distributions, which satisfies the conditions of Lemma~\ref{lem:general-reduction-to-lfht}.
The lemma guarantees an algorithm $A'$ solving $\LFHT(P,Q,s_1,s_2)$ with $s_1 \le T$ samples from $Q$ and $s_2 \le \frac{80T}{k}$ samples from $D$.
From Lemma~\ref{lem:lfht}, such an algorithm must satisfy:
$$\begin{cases}
	s_1^2s_2 \gtrsim (\frac{n}{\varepsilon^2})^2 \quad  s_1s_2 \gtrsim \frac{n}{\varepsilon^4} \quad s_2 \gtrsim \frac{1}{\varepsilon^2}  & \text{if } n \gtrsim \frac{1}{\varepsilon^4}\\
	s_1 \gtrsim \frac{\sqrt{n}}{\varepsilon^2}\quad  s_1s_2 \gtrsim \frac{n}{\varepsilon^4} \quad
	s_2 \gtrsim \frac{1}{\varepsilon^2} & \text{if } n \lesssim \frac{1}{\varepsilon^4}
\end{cases}$$
Substituting $s_1 \le T$ and $s_2 \le \frac{80T}{k}$:
\paragraph{Case 1: $n \lesssim 1/\varepsilon^4$.}
The analysis is identical to Part 1, yielding $T \gtrsim \frac{k+\sqrt{nk}}{\varepsilon^2}$.

\paragraph{Case 2: $n \gtrsim 1/\varepsilon^4$.}
\begin{itemize}
	\item $\frac{80T}{k} \gtrsim s_2 \gtrsim \frac{1}{\varepsilon^2} \implies T \gtrsim \frac{k}{\varepsilon^2}$
	\item $T (\frac{80T}{k}) \gtrsim s_1s_2 \gtrsim \frac{n}{\varepsilon^4} \implies T \gtrsim \frac{\sqrt{nk}}{\varepsilon^2}$
	\item $T^2 (\frac{80T}{k}) \gtrsim s_1^2 s_2 \gtrsim \frac{n^2}{\varepsilon^4} \implies T \gtrsim \left(\frac{n^2k}{\varepsilon^4}\right)^{1/3}$
\end{itemize}

In the $n \lesssim \frac{1}{\varepsilon^4}$ regime we have $ \frac{\sqrt{nk}}{\varepsilon^2} \gtrsim \left(\frac{n^2k}{\varepsilon^4}\right)^{1/3}$, so we merge the two cases to arrive at the lower bound: $\Omega\left(\left(\frac{n^{2}k}{\varepsilon^{4}} \right)^{1/3}+ \frac{k+\sqrt{nk}}{\varepsilon^{2}}\right)$.

\subsubsection{Proof of Lemma \ref{lem:general-reduction-to-lfht}}
\begin{proof}
Let $A$ be the algorithm as stated in Theorem~\ref{thm:combined-lb-lfht}. 
If the $k$ distributions given as input to $A$ cannot be partitioned into $\{\mathcal{I}, [k]\setminus \mathcal{I}\}$ with $|\mathcal{I}| = r$, then the behaviour of $A$ could in principle be arbitrary; for example, it may attempt to draw more than $T$ samples and/or it may not return a partition of size $(r,k-r)$. However, $A$ can be trivially modified to always terminate after $T$ samples and return a partition of size $(r,k-r)$. Let  $\mathcal{I}_A$ denote the subset returned by $A$ of size $k-r$.

Note that the problem is symmetric with respect to the cluster sizes $r$ and $k-r$. 
    Therefore, without loss of generality, we assume in this proof that $r \ge k/2$.
	
	\begin{lemma}\label{lem:typical_dist_general}
		Consider the following random input instance for $A$: let $(P,Q) \sim \Lambda_{\mathrm{alt}}$.
        Set $D_i = P$  for all $i \in \{r+2,\dots,k\}$ and $D_i = Q$ for all $i \in \{1,\dots,r+1\}$.
       Let the number of samples from $D_j$ be $N_j$ for all $j \in [k]$. Then there exists an index $j \in \{1,\dots,r+1\}$ (determinable given $A$ and $\Lambda_{\mathrm{alt}}$) such that:
		\begin{itemize}
			\item $\Pr[\mathcal{I}_A = \{r+2,\dots,k\} \cup \{j\}] \le \frac{4}{k}$
			\item $\mathbb{E}[N_j] \le \frac{4T}{k}$
		\end{itemize}
	\end{lemma}
	\begin{proof}
		Let $S_i = \{i\} \cup \{r+2,\dots,k\}$. Since $\sum_{i \in [r+1]} \Pr[\mathcal{I}_A = S_i] \le 1$, by an averaging argument, there are more than $(r+1)/2$ indices $i$ for which $\Pr[\mathcal{I}_A = S_i] \le 2/(r+1)$.
		Furthermore, $\sum_{i \in [r+1]} \mathbb{E}[N_i] \le T$, so there are more than $(r+1)/2$ indices $i$ for which $\mathbb{E}[N_i] \le 2T/(r+1)$.
		Thus, there exists at least one $j \in [r+1]$ in both sets. For this $j$, since $r \ge k/2$ (implying $r+1 > k/2$), we have:
		$\Pr[\mathcal{I}_A = S_j] \le \frac{2}{r+1} < \frac{4}{k}$
		$\mathbb{E}[N_j] \le \frac{2T}{r+1} < \frac{4T}{k} $.
	\end{proof}
	
	We now present algorithm $A'$ to solve $\LFHT(P,Q,T,\frac{80T}{k})$ when  $(P,Q) \sim \Lambda_{\mathrm{alt}}$. $A'$ has sample access to $P$, $Q$, and an unknown $D \in \{P, Q\}$. 
	\begin{enumerate}
		\item $A'$ selects the index $j$ from Lemma~\ref{lem:typical_dist_general}.
		\item $A'$ sets up an instance for $A$:
		\begin{itemize}
			\item $D_j$ is set to $D$.
			\item Distributions $\{1,\dots,r+1\} \setminus \{j\}$ are set to $Q$.
			\item Distributions $\{r+2,\dots,k\}$ are set to $P$.
		\end{itemize}
		\item $A'$ calls $A$. When $A$ requests a sample from $D_i$, $A'$ queries the corresponding oracle $D$, $P$, or $Q$. That is, if $A$ requests a sample from $D_i$ then  $A'$ queries $P$ when $i \in \{r+2,\dots,k\}$, $Q$ when $i \in \{1,\dots,r+1\}\setminus \{j\}$, and $D$ when $i =j$.
		\item If $N_j$ (number of samples from $D$) exceeds $\frac{80T}{k}$, $A'$ aborts.
		\item Let $S_j = \{j\} \cup \{r+2,\dots,k\}$. $A'$ returns:
		\begin{itemize}
			\item $Q$ if the run completes ($N_j \le \frac{80T}{k}$) AND $\mathcal{I}_A \neq S_j$.
			\item $P$ if the run aborts ($N_j > \frac{80T}{k}$) OR $\mathcal{I}_A = S_j$.
		\end{itemize}
	\end{enumerate}
	We proceed to analyze the sample complexity and correctness of $A'$.  Clearly $A'$ draws at most $T$ samples from both $P$ and $Q$ and at most $\frac{80T}{k}$ samples from $D$. We now show that $A'$ correctly classifies $D \in \{P,Q\}$ with required probability.
	\paragraph{Case 1: $D = Q$.}
	The instance given to $A$ has $r+1$ copies of $Q$ and $k-r-1$ copies of $P$. This is the setup of Lemma~\ref{lem:typical_dist_general}.
	By the lemma, $\Pr[\mathcal{I}_A = S_j] \le 4/k$.
	By Markov's inequality, $\Pr[N_j > \frac{80T}{k}] \le \mathbb{E}[N_j] / (\frac{80T}{k}) \le (4T/k) / (\frac{80T}{k}) = 1/20$.
	Assuming $k \ge 80$, we have $4/k \le 1/20$.
	By the union bound, the probability that $A'$ fails (i.e., aborts or $\mathcal{I}_A = S_j$) is $\le 1/20 + 1/20 = 0.1$.
	Thus, $A'$ correctly returns $Q$ with probability $\ge 0.9$.
	\paragraph{Case 2: $D = P$.}
	The instance given to $A$ has $r$ copies of $Q$ and $k-r$ copies of $P$. This is an instance $A$ was designed for. The true set of base indices is $S_j = \{j\} \cup \{r+2,\dots,k\}$. $A$ guarantees that $\Pr[\mathcal{I}_A = S_j] \ge 0.9$. In this event, $A'$ correctly returns $P$. Thus, $A'$ is correct with probability $\ge 0.9$.
	
	Hence, we find that $A'$ solves $\LFHT(P, Q,s_1,s_2)$ with probability $\ge 0.9$. The total number of samples $A'$ draws from the $Q$ oracle is $s_1 = \sum_{i \in [r+1]\setminus\{j\}} N_i \le T$. The total samples from $D$ is $s_2 = N_j \le \frac{80T}{k}$ (as it aborts otherwise).
\end{proof}

\section{Tightness of 2-Clustering Bounds}\label{sec:discussion}    
Here we provide a case-by-case comparison of our upper and lower bounds, as summarized in Table~\ref{tab:summary_gaps}. In both the ``One-Known-One-Unknown'' and ``Both-Unknown'' settings: the bounds are tight up to a $\log k$ factor. We now discuss these gaps in detail.

Across all scenarios, the $r$-dependent terms in our sample complexity are tight up to $\log k$ factors (this can be seen directly in Table \ref{tab:summary_gaps}). Hence, whenever the $r$-dependent term dominates, the overall upper bound of the algorithm matches the lower bound (up to a $\log k$ factor). It therefore remains to compare the $r$-independent terms from Stage 2.

\paragraph{One Known, One Unknown}
\begin{itemize}
    \item \textbf{($n \gtrsim k\log(k)$):} Upper bound $O(\frac{\sqrt{nk\log(k)}}{\varepsilon^2})$; Lower bound $\Omega(\frac{\sqrt{nk}}{\varepsilon^2})$; Gap $\sqrt{\log k}$.

    \item \textbf{($n \lesssim k\log(k)$):} Upper bound $O\left(\frac{k\log(k)}{\varepsilon^2}\right)$; Lower bound $\Omega\left(\frac{k}{\varepsilon^2}\right)$; Gap $\log k$.
\end{itemize}

\paragraph{Both Unknown}
\begin{itemize}
    \item \textbf{($n \gtrsim \frac{k\log(k)}{\varepsilon^4}$):} Upper bound $O\left( (\frac{n^2k\log(k)}{\varepsilon^4})^{\frac{1}{3}} \right)$, Lower bound $\Omega\left( (\frac{n^2k}{\varepsilon^4})^{\frac{1}{3}} \right)$; Gap $(\log(k))^{\frac{1}{3}}$.

    \item \textbf{($k\log(k) \lesssim n \lesssim \frac{k\log(k)}{\varepsilon^4}$):} Upper bound $O(\frac{\sqrt{nk\log(k)}}{\varepsilon^2})$; Lower bound $\Omega\left(\frac{\sqrt{nk}}{\varepsilon^2}\right)$; Gap $\sqrt{\log k}$.

    \item \textbf{($n \lesssim k\log(k)$):} Upper bound $O\left(\frac{k\log(k)}{\varepsilon^2}\right)$; Lower bound $\Omega\left(\frac{k}{\varepsilon^2}\right)$; Gap: $\log k$.
\end{itemize}
    \section{Miscellaneous Proofs}
    \subsection{Proof of Lemma~\ref{lem:coupon} (Coupon Collector Type Result)}\label{sec:coupon}
    \coupon*
    \begin{proof}
    Note that sampling without replacement yields a strictly lower failure probability than sampling with replacement; for simplicity, we analyze the latter to establish the upper bound.
    
    Let $C_1, \dots, C_{\du}$ denote the sets of indices corresponding to the $\du$ unknown clusters. By the problem promise, the size of each cluster is at least $r$, i.e., $|C_j| \ge r$ for all $j \in [\du]$. For a single index drawn uniformly at random from $[k]$, the probability that it belongs to cluster $C_j$ is $p_j = \frac{|C_j|}{k} \ge \frac{r}{k}$. Let $E_j$ be the event that the cluster $C_j$ is absent from the sample set $\mathcal{S}$. Since the samples are independent, we have:
    $\Pr[E_j] = (1 - p_j)^{|\mathcal{S}|} \le \left(1 - \frac{r}{k}\right)^{|\mathcal{S}|} \le e^{-|\mathcal{S}|\frac{r}{k}}$ using the fact that $1-x \le e^{-x}$. By the union bound over all $\du$ clusters we find the probability of at least one cluster $C_j$ not being found:
    $$ \Pr\left[\bigcup_{j=1}^{\du} E_j\right] \le \sum_{j=1}^{\du} \Pr[E_j] \le \du e^{-|\mathcal{S}|\frac{r}{k}}. $$
    To ensure this failure probability is at most $\frac{1}{9}$, we set:
    $\du e^{-|\mathcal{S}|\frac{r}{k}} \le \frac{1}{9}$. Thus, choosing $|\mathcal{S}| \geq  \frac{k}{r} \log(9\du)$ guarantees that all unknown clusters are discovered with probability at least $\frac{8}{9}$. 
    \end{proof}
    \subsection{Proof of Lemma~\ref{lem:majorise} (Collision-Type Bounds)}\label{sec:majorise}
    
    In the following, recall that $\mathcal{S}$ represents a random multi-set, and $S = {\rm set}(\mathcal{S})$ is the resulting set with duplicates removed. 
    
    For a vector $\vec{v} = (v_1,v_2,\ldots v_n)$, let $\vec{v}^{\downarrow}$ denote the permutation of $\vec{v}$ that follows a non-increasing order, i.e., $\vec{v}^{\downarrow} = (v^{\downarrow}_1, \ldots, v^{\downarrow}_n)$ is a permutation of $\vec{v}$ such that ${v}^{\downarrow}_1 \geq {v}^{\downarrow}_2 \ldots \geq {v}^{\downarrow}_n$. For vectors $\vec{v}$ and $\vec{w}$ such that $\sum_{i \in [n]} \vec{v}(i) = \sum_{i \in [n]} \vec{w}(i)$, we say that $\vec{v}$ is \emph{majorised} by $\vec{w}$, denoted as $\vec{w} \succcurlyeq \vec{v}$ if $\sum_{i=1}^{ r} w_{i}^{\downarrow} \geq   \sum_{i=1}^{r} v_{i}^{\downarrow}$ for all $r \in [k]$.
    
    We will use the following definition of Schur-Concave functions.
    \begin{definition}[Sec.~3 Def.~A.1 in~\citep{MOA11}]\label{def:schurconcave}
        A real-valued function $f$ is said to be Schur-concave if $\vec{q} \succcurlyeq \vec{p}  \implies  f(\vec{q}) \le f(\vec{p})$.
    \end{definition}
    
    \majorise*
    \begin{proof}
    	Consider the case that $S$ results from $\mathcal{S}$ containing $s_1$ i.i.d.~samples from $P$.  
        We wish to bound
    	\begin{align*}
    		\Pr_{\mathcal{S} \sim P^{\otimes s_1}}[U(S) \ge t] = \Pr_{\mathcal{S} \sim P^{\otimes s_1}}[|S| \ge tn].
    	\end{align*}
    	where we used the fact that $U(S) = \frac{|S|}{n}$.
    	
    	For a distribution $D$, let $\vec{D} = (d_1, \dots, d_n)$ be its probability vector. We define $f(\vec{D}) = \Pr_{\mathcal{S} \sim D^{\otimes s_1}}\left[  \frac{|S|}{n} \geq t \right]$.  It is known that $f$ is Schur-concave for all $n$ and $t$ (Proposition E.11.b in~\citep{MOA11}). Moreover, it is easily verified that the uniform distribution $U$ is such that $\vec{P}\succcurlyeq \vec{U} $, i.e., $\vec{U}$ is majorised by every other distribution $\vec{P}$.  We can therefore apply Definition~\ref{def:schurconcave} as follows:
        \begin{alignat*}{2}
         & & \vec{P} & \succcurlyeq \vec{U} \\
    \implies &\qquad  & f(\vec{P}) & \le f(\vec{U}) \\
    \implies & & \Pr_{\mathcal{S} \sim P^{\otimes s_1}}\left[ \frac{|S|}{n} \geq t \right] & \leq \Pr_{\mathcal{S} \sim U^{\otimes s_1}}\left[ \frac{|S|}{n} \geq t \right] \\
    \implies & & \Pr_{\mathcal{S} \sim P^{\otimes s_1}}[U(S) \ge t] & \leq \Pr_{\mathcal{S} \sim U^{\otimes s_1}}[U(S) \ge t].
    \end{alignat*}
    \end{proof}
    
    \section{Adapting to Unknown Cluster Sizes} \label{sec:proofofadaptivity}
    
    Our algorithms, as presented, require knowledge of the parameter $r$ representing the size of one of the clusters. This parameter is used only in Stage 1 to determine the number of samples needed to find an exemplar from each cluster; Stage 2 does not depend on $r$.  In this section, we describe a variation that does not require knowledge of $r$, but requires more rounds of adaptivity, namely, $O\big(\log\frac{k}{r}\big)$ instead of two.
    
    We focus here on the case that one distribution is known to be uniform and the other is unknown; the argument for `both unknown' is entirely analogous and thus omitted.  Recall from Section \ref{sec:1known1unknown} that Stage 1 selects a subset of $\lceil \frac{18k}{r} \rceil$ distributions uniformly at random, and performs uniformity testing with a suitably-chosen target success probability.  We showed in the analysis of Stage 1 that the ``good event''  $\overline{E}_1 \wedge \overline{E}_2$ occurs with probability at least $\frac{8}{9}$, where $\overline{E}_1$ ensures that at least one non-uniform distribution appears in the subset, and $\overline{E}_2$ ensures that all of the uniformity tests are correct.
    
    When $r$ is unknown, we modify this procedure to \emph{iteratively add and test distributions until the first success}.  We do this in rounds indexed by $i=1,2,\dotsc$, with the $i$-th round testing an additional $2^i$ distributions for uniformity, each with an associated error probability of $c\frac{2^{-i}}{i^2}$ for some constant $c$.  As shown in the known-$r$ case, with probability at least $\frac{17}{18}$, one of the first $\lceil \frac{18k}{r} \rceil$ distributions considered will be non-uniform (event $\overline{E}_1$ above), so our main goal is to obtain a counterpart to event $\overline{E}_2$.  
    
    Let $i^*$ be the iteration at which the number of distributions becomes at least $\lceil \frac{18k}{r} \rceil$ (and thus at most $2\lceil \frac{18k}{r} \rceil$ since we are doubling in each iteration).  
    By the above choice of target error probability and the union bound, the probability of receiving an erroneous decision in any of these uniformity tests is at most $\sum_{i=1}^{i^*} c\frac{2^{-i}}{i^2} 2^i \le c \sum_{i=1}^{\infty} \frac{1}{i^2}$, which is at most $\frac{1}{18}$ when $c$ is sufficiently small.  This is the desired counterpart to $\overline{E}_2$, as combined with the good event from the previous paragraph, it ensures the success of Stage 1.
    
    It remains to study the sample complexity.  We first observe that stopping by iteration $i^*$ implies considering at most twice as many distributions (in Stage 1) compared to the known-$r$ case.  
    For known $r$, the union bound incurred a multiplicative $O\big(\log\frac{k}{r}\big)$ factor, whereas here the multiplicative factor becomes $O\big(\log\frac{1}{2^{-i^*}/(i^*)^2}\big) = O(i^*)$.  By the definition of $i^*$ and the fact that the number of distributions added is exponentially increasing (hence dominated by the final term), this simplifies to $O\big(\log\frac{k}{r}\big)$, thus matching the known-$r$ case to within constant factors.  Note also that $i^*$ represents the number of stages of adaptivity by definition.

\section{Adapting to Unknown $\varepsilon$}\label{sec:epsilonadaptivity}
Our algorithms, as presented, require knowledge of the parameter $\varepsilon^* = \tv(P,Q)$ that determines the separation of the two clusters. In this section, we relax the assumption that $\varepsilon^*$ is provided as input. We employ an iterative guess-and-verify strategy: at iteration $j \in \{0,1,2, \ldots\}$, we guess that the separation is $\varepsilon_j = 2^{-j}$ and cluster the distributions using this guess. To do so, we use a clustering algorithm, denoted by $\mathcal{C}(\eta, \delta)$, that returns a correct partition with probability at least $1-\delta$, provided the input distributions are separated by at least $\eta$. We then verify the result using a \textit{tolerant test} (defined below) to determine if the separation is indeed at least $\varepsilon_j$.  See Algorithm~\ref{alg:epsadaptivity} for the details. 

The sample complexity of clustering algorithms (for known $\varepsilon$) typically depends on the domain size $n$, the number of distributions $k$, and the cluster size $r$, but we will suppress these dependencies in order to reduce notation. For our adaptive procedure, we will henceforth assume a sample complexity upper bound $N_{\mathcal{C}}(\varepsilon, \delta)$ that scales polynomially with $1/\varepsilon$ and logarithmically with $1/\delta$, as is the case for the algorithms presented in the paper. Thus, we have $N_{\mathcal{C}}(\varepsilon,\delta) \in O(\text{poly}(1/\varepsilon) \cdot \log(1/\delta))$.  

\begin{lemma}{\citep[Theorem 4]{VV11}}
    Given sample access to two distributions $P$ and $Q$, parameters $\delta \in (0,1)$, and a threshold $0 \le \varepsilon \le 1$, a tolerant test $\mathcal{T}(P,Q,\varepsilon, \delta)$ is an algorithm that with probability at least $1-\delta$:
    \begin{itemize}
        \item returns $\mathsf{NEAR}$ if $\tv(P,Q) \le \frac{\varepsilon}{2}$;
        \item returns $\mathsf{FAR}$ if $\tv(P,Q) \ge \varepsilon$.
    \end{itemize}
    If $\varepsilon/2 < \tv(P,Q) < \varepsilon$, the test may return either $\mathsf{NEAR}$ or $\mathsf{FAR}$. Then, the sample complexity of tolerant testing satisfies an upper bound of the form:
    \begin{equation}
        N_{\mathcal{T}}(n,\varepsilon,\delta) = O\left(\frac{n}{\log(n)\varepsilon^2}\log\frac{1}{\delta}\right)\label{eq:tolerant_n}
    \end{equation}
    \end{lemma}

The following theorem states the sample complexity of our adaptive 2-clustering algorithm in terms of the sample complexities of the underlying non-adaptive clustering and tolerant testing algorithms.

\begin{algorithm2e}[t]
    \caption{2-Clustering Adaptive to Unknown $\epsilon$}
    \label{alg:epsadaptivity}
    
    \SetKwInput{KwOracles}{Oracles}
    
    \KwIn{Sample access to $\{D_i\}_{i=1}^k$, failure probability $\delta$.}
    \KwOracles{Clustering algorithm $\mathcal{C}$, tolerant tester $\mathcal{T}$.}
    \KwOut{Partition $\{\mathcal{I}_1, \mathcal{I}_2\}$.}
    
    \For{$j = 0, 1, \dots$}{
        $\varepsilon_j \gets 2^{-j}$\;
        $\delta_j \gets \frac{2\delta}{\pi^2 (j+1)^2}$\;
        $\{\mathcal{I}_1, \mathcal{I}_2\} \gets \mathcal{C}(\varepsilon_j, \delta_j)$\;
        Pick arbitrary indices $a \in \mathcal{I}_1$ and $b \in \mathcal{I}_2$\;
        \If{$\mathcal{T}(D_a, D_b, \varepsilon_j, \delta_j)$ returns $\mathsf{FAR}$}{
            $\{\mathcal{I}_1, \mathcal{I}_2\} \gets \mathcal{C}(\frac{\varepsilon_j}{2}, \delta_j)$\;
            \Return $\{\mathcal{I}_1, \mathcal{I}_2\}$\;
        }
    }
\end{algorithm2e}
\begin{theorem}\label{thm:geomsearch}
    Let $N_{\mathcal{C}}(\varepsilon, \delta)$ (resp., $N_{\mathcal{T}}(n,\varepsilon,\delta)$) be any sample complexity upper bounds for clustering (resp., tolerating testing) whose $(\epsilon,\delta)$ dependence scales as $O\big(\textrm{poly}(1/\varepsilon) \cdot \log(1/\delta)\big)$.  
    Then, Algorithm~\ref{alg:epsadaptivity} returns the true partition with probability at least $1-\delta$, with sample complexity at most
    $$ O\left( \left( N_{\mathcal{C}}(\varepsilon^*, \delta) + N_{\mathcal{T}}(n, \varepsilon^*, \delta) \right) \left[ 1 + \frac{\log\log(1/\varepsilon^*)}{\log(1/\delta)} \right] \right), $$
    where $\varepsilon^*$ is the true separation.
\end{theorem}
\begin{remark}
    We note that from \eqref{eq:tolerant_n} the tolerant testing term $N_{\mathcal{T}}$ scales linearly with the domain size $n$ (up to log factors) but is independent of the number of distributions $k$. In contrast, the clustering complexity $N_{\mathcal{C}}$ grows with both $k$ and $n$ (e.g., $\sqrt{nk}$), as seen in Table \ref{tab:summary_gaps}. Thus, in regimes where $k$ is large, the cost is dominated by $N_{\mathcal{C}}$, and the overhead from the tolerant testing verification is negligible.  As a concrete example, if $k = \Omega(n)$ then the $N_{\mathcal{T}}$ does not change the overall scaling, since $N_{\mathcal{C}}$ is always at least as high as $\frac{k}{\epsilon^2}$.  
    However, for certain regimes of smaller $k$, the tightness of our upper bound remains open.
\end{remark}

\begin{proof}
The probability that the clustering algorithm or the tolerant tester errs in iteration $j$ is bounded by $\delta_j$. Since in each iteration we make no more than $3$ calls to the oracles, the total error is bounded by $3 \delta_j$. Using the identity $\sum_{j=0}^{\infty} (j+1)^{-2} = \pi^2/6$, a union bound over all iterations ensures the total error probability is at most $3\sum_{j=0}^{\infty} \frac{2\delta}{\pi^2 (j+1)^2} = 3\delta/3 = \delta$. We henceforth assume that all oracle calls return correct results according to their promises.
    
Let $j^*$ be the first index such that $\varepsilon_{j^*} \le \varepsilon^*$; since $\varepsilon_j = 2^{-j}$, we have $j^* = \lceil \log(\frac{1}{\varepsilon^*})\rceil$. For any $j < j^*$, the tolerant tester is invoked with threshold $\varepsilon_j$. If $\varepsilon^* \le \frac{\varepsilon_j}{2}$, then $\mathcal{T}$ correctly returns $\mathsf{NEAR}$, whereas if $\frac{\varepsilon_j}{2} < \varepsilon^* < \varepsilon_j$ (the ``undefined'' region), then $\mathcal{T}$ may return $\mathsf{FAR}$ or $\mathsf{NEAR}$.  Then:
\begin{itemize}
    \item If $\mathcal{T}$ returns $\mathsf{NEAR}$, the loop continues.
    \item If $\mathcal{T}$ returns $\mathsf{FAR}$, the algorithm executes Line 10, calling $\mathcal{C}(\frac{\varepsilon_j}{2}, \delta_j)$. Since this can only occur in the region where $\varepsilon^* > \frac{\varepsilon_j}{2}$, the separation condition for $\mathcal{C}$ is satisfied ($\tv(P,Q) \ge \frac{\varepsilon_j}{2}$), and it returns the correct partition.
\end{itemize}
At iteration $j=j^*$, we have $\varepsilon_{j^*} \le \varepsilon^* < 2\varepsilon_{j^*}$. The clustering algorithm $\mathcal{C}(\varepsilon_{j^*}, 
\delta_{j^*})$ succeeds since $\varepsilon^* \ge \varepsilon_{j^*}$. The tolerant tester verifies $\tv(D_a, D_b) = \varepsilon^* \ge \varepsilon_{j^*}$ and returns $\mathsf{FAR}$. The subsequent ``safety call'' $\mathcal{C}(\varepsilon_{j^*}/2, 
\delta_{j^*})$ also succeeds since $\varepsilon^* \ge \varepsilon_{j^*} > \varepsilon_{j^*}/2$.

The total sample complexity is bounded by the sum of sample costs across all iterations. Since the complexities $N_{\mathcal{C}}$ and $N_{\mathcal{T}}$ scale polynomially with $1/\varepsilon$ (as $O(\varepsilon^{-p})$ for some constant $p > 0$), the sum forms a geometric series dominated by its final term. Specifically,
\begin{align*}
    &\sum_{j=0}^{j^*} \left( N_{\mathcal{C}}(\varepsilon_j, \delta_j) + N_{\mathcal{T}}(n,\varepsilon_j, \delta_j) \right)  \le O\left( N_{\mathcal{C}}(\varepsilon_{j^*}, \delta_{j^*}) + N_{\mathcal{T}}(n,\varepsilon_{j^*}, \delta_{j^*}) \right).
\end{align*}

By our choices of $\varepsilon_j$, $\delta_j$, and $j^*$, we have $\varepsilon_{j^*} = \Theta(\varepsilon^*)$ and $\delta_{j^*} = \Theta\left(\frac{\delta}{\log^2(1/\varepsilon^*)}\right)$. Moreover, by assumption, the sample complexities factor as $N(\varepsilon, \delta) = f(\varepsilon) \cdot \log(1/\delta)$. Substituting $\delta_{j^*}$ yields:
    $$ \log\left(\frac{1}{\delta_{j^*}}\right) = \log\left( \Theta\left( \frac{\log^2(1/\varepsilon^*)}{\delta} \right) \right) = \log\left(\frac{1}{\delta}\right) + 2\log\log\left(\frac{1}{\varepsilon^*}\right) + O(1). $$
    Therefore, the complexity $N(\varepsilon_{j^*}, \delta_{j^*})$ decomposes into the standard non-adaptive cost plus an additive overhead:
    $$ N(\varepsilon_{j^*}, \delta_{j^*}) = f(\varepsilon^*) \log\left(\frac{1}{\delta}\right) + f(\varepsilon^*) \cdot O\left( \log\log\left(\frac{1}{\varepsilon^*}\right) \right). $$
    Applying this to both $\mathcal{C}$ and $\mathcal{T}$, the total sample complexity is bounded by
    $$ O\left( \left( N_{\mathcal{C}}(\varepsilon^*, \delta) + N_{\mathcal{T}}(n, \varepsilon^*, \delta) \right) \left[ 1 + \frac{\log\log(1/\varepsilon^*)}{\log(1/\delta)} \right] \right).$$
\end{proof}

%% file: extension.tex
\section{Details and Proofs for $d$-Clustering}\label{sec:extension}

In this appendix, we extend our results to $d$-clustering distributions where $d > 2$. We generalize the problem setup to allow for a mix of known and unknown cluster centers. Specifically, we consider a setting with $d$ total clusters, comprising $\dk$ unknown centers and $\du$ known centers, such that $d = \du + \dk$.

\begin{framed}
	\noindent\textbf{The Distribution $d$-Clustering Problem}
	\begin{description}[leftmargin=1em]
		\item[Input:] Sample access to $k$ distributions $\{D_i\}_{i=1}^k$ over domain $[n]$, parameter $\varepsilon \in (0,1]$, and a set of known reference distributions $\mathcal{Q} = \{Q_1, \dots, Q_{\dk}\}$. The distributions $\{D_i\}_{i=1}^k$ are allowed to be queried adaptively.
		\item[Promise:] There exists a hidden partition $\{\mathcal{I}_j\}_{j=1}^{d}$ of $[k]$ that satisfies the following:
		\begin{itemize}[leftmargin=1.5em]
			\item The clusters correspond to $d$ centers, comprised of the $\dk$ known distributions and $\du$ unknown distributions (where $d = \du + \dk$).
			\item $\forall{(a, b \in \mathcal{I}_j)}$,  $D_a = D_b$.
			\item $\forall{(a\in \mathcal{I}_i,b \in \mathcal{I}_j)}$ with $i \neq j$, $d_{\mathrm{TV}}(D_a,D_b) \ge \varepsilon$.
		\end{itemize}
		\item[Goal:] With probability at least $2/3$, output the true partition $\{\mathcal{I}_j\}_{j=1}^d$.
	\end{description}
\end{framed}

To solve this problem, we propose a two-stage algorithm. In Stage 1, we identify $\du$ exemplars, one for each unknown distribution. In Stage 2, we classify every distribution in the input set against all $d$ exemplars (both known and discovered) to recover the partition.

\subsection{Stage 1: Finding All Exemplars}
Since $\dk$ centers are already known, our goal is to obtain sample access to the remaining $\du$ unknown centers. We achieve this by sampling a subset of distributions uniformly at random. Let $r$ be the size of the smallest cluster among those with unknown distributions. The problem of finding at least one member from each of the $\du$ unknown clusters is analogous to the Coupon Collector's Problem. We state this formally as follows and defer the proof to Appendix~\ref{sec:coupon}.
\begin{restatable}{lemma}{coupon}\label{lem:coupon}
    Let $\mathcal{K} \subseteq [k]$ be a subset of indices chosen uniformly at random without replacement. If $|\mathcal{K}| = \lceil \frac{k}{r} \log(9\du)\rceil$, then with probability at least $\frac{8}{9}$, the set $\{D_i\}_{i \in \mathcal{K}}$ contains at least one representative from each of the $\du$ unknown clusters.
\end{restatable}
Once the subset $\mathcal{K}$ of size $\lceil \frac{k}{r} \log(9\du)\rceil$ is drawn, we must verify which distributions are new exemplars. We first filter out the distributions that correspond to the known exemplars (if $\dk \geq 1$), and hence cannot be the new unknown exemplars. To do this, we perform the following procedure: for each $D \in \mathcal{K}$, we test if $D$ is equivalent to any distribution in $\mathcal{Q}$ (known centers) and if so, we remove the said distribution from $\mathcal{K}$. This requires running $O(|\mathcal{K}| \cdot \dk)$ standard identity tests against each distribution in $\mathcal{Q}$, drawing $O\left(  \frac{|\mathcal{K}|\sqrt{n}}{\varepsilon^2} \log_+\left(\dk\right)\right)$\footnote{We use $\log_+(x)$ to denote $\max(\log(x),1)$.} samples in total (see Table~\ref{tab:dist_testing}). The $\log$ factor arises as we reuse the samples from each distribution $\dk$ times, hence each test must succeed with probability $1-O(1/\dk)$. 

Then, we apply the Find-All-Exemplars algorithm (described in the following subsection) to identify the unique set of unknown exemplars $\mathcal{P} = \{P_1, \dots, P_{\du}\}$. Since the filtered subset has at most $|\mathcal{K}|$ distributions, running Find-All-Exemplars has the following sample complexity (see Theorem \ref{thm:main_equiv_split_test_extended} below):
$$O\left(\max\left(\left(\frac{\du |\mathcal{K}|^2 n^2}{\varepsilon^4}\right)^{1/3}, \frac{|\mathcal{K}|\sqrt{n}}{\varepsilon^2}\right)\log\left(|\mathcal{K}|\right)\right).$$

Combining the complexity of the initial filtering operation and the subsequent run of Find-All-Exemplars, and accounting for the fact that $\du = \dk = O(1)$ we see that the sample complexity of Stage 1 is:
$$O\left(\max\left(\left(\frac{ k^2 n^2}{r^2\varepsilon^4}\right)^{1/3}, \frac{k\sqrt{n}}{r\varepsilon^2}\right)\log\left(\frac{k}{r}\right)\right).$$

If the minimum cluster size $r$ is unknown, we can employ the adaptive strategy described in Appendix~\ref{sec:proofofadaptivity}. Specifically, we proceed in rounds, geometrically increasing the size of the random subset $\mathcal{K}$ and running Algorithm~\ref{alg:uneqequitestextended} in each round, terminating once all $\du$ distinct unknown exemplars are identified.

\subsubsection{The Find-All-Exemplars Algorithm}\label{sec:findallexemplars}
The procedure for finding one exemplar from each cluster is given in Algorithm \ref{alg:uneqequitestextended}.  Since we  apply this procedure on a suitably-chosen subset of our distributions rather than the full set of distributions, we use generic notation $m$ for the number of distributions here.
\begin{algorithm2e}[h!]
	\caption{Find-All-Exemplars}
	\label{alg:uneqequitestextended}
	\DontPrintSemicolon
	\LinesNumbered
	\KwIn{$\varepsilon \in (0,1)$, distributions $\{D_i\}_{i=1}^{m}$, $\du$ (putative number of clusters).}
	\KwOut{A set of indices $\mathcal{E} \subseteq [m]$ representing one exemplar from each cluster, or a report that fewer than expected clusters found.}
	
	$s_1 \gets \min(n,(\frac{nm}{\du\varepsilon^{2}})^{2/3})$\;
	$\delta' \gets \frac{1}{9m^2}$\; 
	$b \gets \sqrt{\frac{9\du}{s_1}}$\;
	$s_2 \gets Cn s_1^{-1/2} \frac{\log(1/{\delta'})}{\varepsilon^2}$ with $C$ being the hidden $O(\cdot)$ constant in Lemma~\ref{lem:L2test}\;
	
	Draw $s_2$ samples from every distribution $D_i$ for $i \in \mathcal{U}$\;
	$\mathcal{E} \gets \emptyset$\; 
	\While{$\mathcal{U} \neq \emptyset$}{

		Pick an arbitrary index $j \in \mathcal{U}$ to be the exemplar\;
		$\mathcal{U} \gets \mathcal{U} \setminus \{j\}$\;
		$\mathcal{E} \gets \mathcal{E} \cup \{j\}$\; 
		\If{$|\mathcal{E}|= \du$}{ \Return $\mathcal{E}$}
			Draw $s \sim \mathrm{Poi}(s_1)$ samples from $D_j$ to form sketch $\mathcal{S}$\;
			\If{$s > e^2\log(9\du )s_1$}{ \Return $\bot$ }
			$\mathcal{C} \gets \emptyset$\;
			\For{$i \in \mathcal{U}$}{
				Run the test from Lemma~\ref{lem:L2test} comparing $D_{i,\mathcal{S}}$ and $D_{j,\mathcal{S}}$ with the pre-drawn $s_2$ samples, and with parameters $\varepsilon$, $\delta'$ and $b$ \label{line:test}\;
				\If{test returns $\tv(D_{i,\mathcal{S}},D_{j,\mathcal{S}}) = 0$}{
					$\mathcal{C} \gets \mathcal{C} \cup \{i\}$\;
				}
			}
		$\mathcal{U} \gets \mathcal{U} \setminus 
			\mathcal{C}$\;}
	
	\If{$|\mathcal{E}| < \du$}{
		\Return \text{Fewer than expected clusters found } 
	}
\end{algorithm2e}

We will prove the following theorem, which mostly uses the same ideas as Theorem~\ref{thm:main_equiv_split_test}.

\begin{theorem}\label{thm:main_equiv_split_test_extended}
	Find-All-Exemplars (Alg.~\ref{alg:uneqequitestextended}) takes in parameters $\varepsilon \in (0,1)$ and $m$ distributions containing at most $\du$ unique clusters. With probability at least $2/3$:
	\begin{enumerate}
		\item If there are exactly $\du$ clusters, it returns a set $\mathcal{E}$ containing exactly one index from each cluster.
		\item If there are fewer than $\du$ clusters, it returns $\bot$.
	\end{enumerate}
	In the case of constant $\du$, the sample complexity is:
	$$O\left(\max\left(\left(\frac{m^2 n^2}{\varepsilon^4}\right)^{1/3}, \frac{m\sqrt{n}}{\varepsilon^2}\right)\log(m)\right).$$
\end{theorem}

\begin{proof}
In each iteration of the while loop, the algorithm picks an arbitrary distribution $D_j$ among those still under consideration, and designates it as an exemplar, adding it to the exemplar set $\mathcal{E}$. Then, the algorithm draws $s$ samples from $D_j$ to create the multiset $\mathcal{S}$, where $s$ is a random variable drawn from a Poisson distribution with mean $s_1$. If $s > c s_1$, where $c = e^2\log(9\du)$, the algorithm terminates immediately and returns $\bot$. The algorithm iterates through the outer while loop at most $\du$ times (once per unique cluster). Let $s^{(\ell)}$ denote the number of samples drawn for the $\ell$-th exemplar, where $s^{(\ell)} \sim \mathrm{Poi}(s_1)$. 
	
Let $E_1$ denote the event that for {any} of the $\du$ iterations, $s^{(\ell)} > c s_1$. By the Chernoff bound, for any $s_1 \ge 1$, we have
$$\Pr[s^{(\ell)} > c s_1] \le \frac{e^{-s_1}(e s_1)^{c s_1}}{(c s_1)^{c s_1}} = \left(\frac{e^{c-1}}{c^c}\right)^{s_1} \le \frac{e^{c}}{c^c} \le \frac{1}{e^c} \leq \frac{1}{9\du},$$ and by a union bound over $\du$ iterations, $\Pr[E_1] \le \frac{1}{9}$.

As discussed in Theorem~\ref{thm:main_equiv_split_test}, the multi-set $\mathcal{S}$ drawn from $D_j$ is used to transform all the distributions into  ``flattened" distributions and the equivalence tests are conducted on these flattened distributions. The $\ell_2$ norm  of these flattened distributions determines the sample complexity of the test, and we know that it is small  in expectation, i.e., $\mathbb{E}[\|D_{j,\mathcal{S}}\|^2_2] \leq \frac{1}{s_1}$.
	
Let $E_{2,\ell}$ denote the event that for exemplar $\ell$ (of the $\du$ exemplars), it is the case that $\|D_{j,\mathcal{S}^{(\ell)}}\|_2 > \sqrt{\frac{9\du}{s_1}}$. By Markov's inequality: 
\begin{align}
\Pr[E_{2,\ell}] = \Pr\left[\|D_{j,\mathcal{S}^{(\ell)}}\|_2 > \sqrt{\frac{9\du}{s_1}}\right] = \Pr\left[\|D_{j,\mathcal{S}^{(\ell)}}\|_2^2 > \frac{9\du}{s_1}\right] \le \frac{\mathbb{E}[\|D_{j,\mathcal{S}^{(\ell)}}\|_2^2]}{9\du/s_1} \le  \frac{1}{9\du}. \label{eqn:markov}
\end{align} 

Let $E_2$ be the event denoting the case that for at least one exemplar, the $||D_{j,\mathcal{S}^{(\ell)}}||_2 > \sqrt{\frac{9\du }{s_1}}$ condition holds. Hence, since $E_2 = \bigcup_{\ell} E_{2,\ell}$, a union bound over $\du$ exemplars gives $\Pr[E_{2}] \le \frac{1}{9}$.

	\paragraph{Testing Event ($E_3$).}
	We henceforth condition on $\overline{E}_1 \cap \overline{E}_2$. On this event, recalling that $b = \sqrt{\frac{9\du}{s_1}}$, the prerequisite condition $b \ge \min (\|D_{i,\mathcal{S}^{(\ell)}}\|_2, \|D_{j,\mathcal{S}^{(\ell)}}\|_2)$ for Lemma~\ref{lem:L2test} is satisfied, since $\min (\|D_{i,\mathcal{S}^{(\ell)}}\|_2, \|D_{j,\mathcal{S}^{(\ell)}}\|_2) \le \|D_{j,\mathcal{S}}^{(\ell)}\|_2 \le b$. The testing subroutine on Line~\ref{line:test} distinguishes between $\tv(D_{i,\mathcal{S}},D_{j,\mathcal{S}}) = 0$ and $\tv(D_{i,\mathcal{S}}, D_{j,\mathcal{S}}) \ge \varepsilon$, which is equivalent to distinguishing between $\tv(D_{i},D_{j}) = 0$ and $\tv(D_{i}, D_{j}) \ge \varepsilon$. The tester draws $s_2 = Cbn\log(1/\delta)/\varepsilon^2 \asymp s_1^{-1/2}n\log(\frac{1}{\delta'})/\varepsilon^2$ samples and errs with probability at most $\delta'$. Let $E_3$ be the event that \emph{any} of the pairwise tests return an incorrect result.
	
	In the worst case, the algorithm compares every candidate against every identified exemplar. The total number of pairwise tests is thus bounded by $m \du \le m^2$. The algorithm sets the error per-test to be $\delta' \gets \frac{1}{9m^2}$, so by a union bound over at most $m^2$ tests, $\Pr[E_3] \le m^2 \delta' \le \frac{1}{9}$.
	
	The total probability of error is $\Pr[E_1] + \Pr[E_2] + \Pr[E_3] \le \frac{1}{3}$.  Conditioned on the success event $\overline{E} = \overline{E}_1 \cap \overline{E}_2 \cap \overline{E}_3$, the algorithm operates as follows: In each iteration, it picks a distribution $j$ from $\mathcal{U}$, which must belong to a previously undiscovered cluster. It correctly identifies all $i \in \mathcal{U}$ equivalent to $j$ and removes them.
	\begin{itemize}
		\item If there are exactly $\du$ clusters, the loop runs $\du$ times, collecting one index per cluster in $\mathcal{E}$.
		\item If there are fewer than $\du$ clusters, the loop terminates with $|\mathcal{E}| < \du$, and the algorithm returns $\bot$.
	\end{itemize}
	
	\paragraph{Sample Complexity.}
	The total sample complexity is determined by the sketching cost paid $\du$ times at most, and the comparison cost (paid for all $m$ distributions): $ O( \du s_1 + m s_2)$. The algorithm chooses $s_1 \asymp \min(n,(\frac{mn}{\du \varepsilon^2})^{2/3})$, and $s_2 \asymp n s_1^{-1/2} \cdot \frac{\log(m)}{\varepsilon^2}$ to balance the costs. This creates two regimes:
	\begin{itemize}
		\item When $\frac{m}{\du} < \varepsilon^2 \sqrt{n}$, we use $s_1 = (\frac{mn}{\du \varepsilon^2})^{2/3}$. The total complexity is $\Theta((\frac{\du m^2 n^2}{\varepsilon^4})^{1/3}\log(m))$.
		\item  $\frac{m}{\du} \ge \varepsilon^2 \sqrt{n}$, we cap the fingerprinting cost at $s_1 = n$.
		The complexity is dominated by the $m$ comparisons: $m s_2 = \Theta(\frac{m\sqrt{n}}{\varepsilon^2}\log(m))$.
	\end{itemize}
	
Combining these regimes, we obtain a sample complexity of
	$$O\left(\max\left(\left(\frac{\du m^2 n^2}{\varepsilon^4}\right)^{1/3}, \frac{m\sqrt{n}}{\varepsilon^2}\right)\log(m)\right),$$
    which simplifies to the stated scaling when $\du$ is constant.
\end{proof}

\subsection{Stage 2: All-Pairs Classification}
In this stage, we have the full set of $d = \du + \dk$ exemplars. Let $\mathcal{E} = \mathcal{Q} \cup \mathcal{P}$ denote this set. To classify the $k$ input distributions, we adopt an \textit{all-pairs} tournament strategy. For every input distribution $D_i$ ($i \in [k]$), and for every distinct pair of exemplars $(E_a, E_b) \in \mathcal{E} \times \mathcal{E}$, we run a specific pairwise test to decide if $D_i$ is closer to $E_a$ or $E_b$:
\begin{enumerate}
	\item {Both Known ($E_a, E_b \in \mathcal{Q}$):} We run \scheffe's test as described in Section~\ref{sec:related}.
	\item {One Known, One Unknown ($E_a \in \mathcal{Q}, E_b \in \mathcal{P}$):} We run \eswt~as used for $d=2$.
	\item {Both Unknown ($E_a, E_b \in \mathcal{P}$):} We run \mlfht~as used for $d=2$.
\end{enumerate}
Specifically, we test $D_i = E_a \text{ vs. } D_i = E_b$. If $D_i$ truly belongs to the cluster of $E_a$, it will win all pairwise comparisons against other exemplars $E_b$.
Since we treat $d$ as a constant, the number of such pairs is $O(1)$. Consequently, the analysis of the sample complexity mirrors the $d=2$ case exactly, with the total complexity dominated by the hardest pairwise distinction summed over $k$ distributions.

\begin{theorem}\label{thm:hybrid_classification}
	Given $\dk$ known distributions and sample access to $\du$ unknown distributions, there exists an algorithm that, with success probability at least $\frac{8}{9}$, classifies $k$ input distributions against the $d = \dk + \du$ exemplars. Assuming $d$ is a constant:
    \begin{itemize}
        \item \textbf{If $\du = 0$:} The sample complexity is $O\left(\frac{k\log k}{\varepsilon^2}\right)$.
        \item \textbf{If $\du = 1$:} The sample complexity matches the bound for the ``One Known, One Unknown'' classification stage (Theorem~\ref{thm:stage2_complexity_one_unknown}).
        \item \textbf{If $\du \ge 2$:} The sample complexity matches the bound for the ``Both Unknown'' classification stage (Theorem~\ref{thm:multilfht}).
    \end{itemize}
\end{theorem}
\begin{proof}
    The algorithm performs $O(d^2 k)$ pairwise tests. Since $d=O(1)$, this scales as $O(k)$. The necessary sample complexity is dictated by the difficulty of the specific pairwise comparisons involved.
    \begin{itemize}
        \item \textbf{Case $\du=0$:} We only compare against known distributions using \scheffe's test, requiring $O(\frac{\log k}{\varepsilon^2})$ samples per distribution.
        \item \textbf{Case $\du=1$:} The hardest comparison is between the single unknown exemplar and the known exemplars. The optimization of samples ($s_1$ from the unknown exemplar, $s_2$ from the inputs) is identical to the derivation in Section \ref{sec:stage2oneunknown}, yielding the bounds in Theorem \ref{thm:stage2_complexity_one_unknown}.
        \item \textbf{Case $\du \ge 2$:} The hardest comparison involves distinguishing between two unknown exemplars. This necessitates the use of \mlfht{} as derived in Section \ref{sec:multilfht}. The constraints and resulting regimes are identical to the classification stage of the ``Both Unknown'' case, yielding the bounds in Theorem \ref{thm:multilfht}.
    \end{itemize}
    We omit the detailed calculations for general $d$ as they follow the identical structure of the proofs in Section \ref{sec:upperbounds} up to constant factors depending on $d$.
\end{proof}

%% file: main.bbl
\begin{thebibliography}{23}
\providecommand{\natexlab}[1]{#1}
\providecommand{\url}[1]{\texttt{#1}}
\expandafter\ifx\csname urlstyle\endcsname\relax
  \providecommand{\doi}[1]{doi: #1}\else
  \providecommand{\doi}{doi: \begingroup \urlstyle{rm}\Url}\fi

\bibitem[Acharya et~al.(2022)Acharya, Canonne, Liu, Sun, and Tyagi]{ACLS+22}
Jayadev Acharya, Clément~L. Canonne, Yuhan Liu, Ziteng Sun, and Himanshu
  Tyagi.
\newblock Interactive inference under information constraints.
\newblock \emph{IEEE Transactions on Information Theory}, 68:\penalty0
  502--516, 2022.

\bibitem[Batu et~al.(2000)Batu, Fortnow, Rubinfeld, Smith, and White]{BFRS00}
Tugkan Batu, Lance Fortnow, Ronitt Rubinfeld, Warren~D Smith, and Patrick
  White.
\newblock Testing that distributions are close.
\newblock In \emph{Symposium on Foundations of Computer Science (FOCS)}, pages
  259--269. IEEE, 2000.

\bibitem[Bhattacharya and Valiant(2015)]{BV15}
Bhaswar Bhattacharya and Gregory Valiant.
\newblock Testing closeness with unequal sized samples.
\newblock \emph{Advances in Neural Information Processing Systems (NeurIPS)},
  28, 2015.

\bibitem[Canonne(2020)]{C20}
Cl{\'{e}}ment~L. Canonne.
\newblock \emph{A Survey on Distribution Testing: Your Data is Big. But is it
  Blue?}
\newblock Graduate Surveys. Theory of Computing Library, 2020.

\bibitem[Canonne(2022)]{C22}
Clément~L. Canonne.
\newblock Topics and techniques in distribution testing: A biased but
  representative sample.
\newblock \emph{Foundations and Trends® in Communications and Information
  Theory}, 19, 2022.

\bibitem[Chan et~al.(2014)Chan, Diakonikolas, Valiant, and Valiant]{CDVV14}
Siu-On Chan, Ilias Diakonikolas, Paul Valiant, and Gregory Valiant.
\newblock Optimal algorithms for testing closeness of discrete distributions.
\newblock In \emph{ACM-SIAM Symposium on Discrete Algorithms (SODA)}, pages
  1193--1203. SIAM, 2014.

\bibitem[Devroye and Lugosi(2001)]{DL01}
Luc Devroye and G{\'a}bor Lugosi.
\newblock \emph{Combinatorial methods in density estimation}.
\newblock Springer Science \& Business Media, 2001.

\bibitem[Diakonikolas and Kane(2016)]{DK16}
Ilias Diakonikolas and Daniel~M Kane.
\newblock A new approach for testing properties of discrete distributions.
\newblock In \emph{Symposium on Foundations of Computer Science (FOCS)}, 2016.

\bibitem[Gerber et~al.(2023)Gerber, Han, and Polyanskiy]{GHP23}
Patrik~R Gerber, Yanjun Han, and Yury Polyanskiy.
\newblock Minimax optimal testing by classification.
\newblock In \emph{Conference on Learning Theory (COLT)}, pages 5395--5432.
  PMLR, 2023.

\bibitem[Gerber and Polyanskiy(2024)]{GP24}
Patrik~R{\'o}bert Gerber and Yury Polyanskiy.
\newblock Likelihood-free hypothesis testing.
\newblock \emph{IEEE Transactions on Information Theory}, 2024.

\bibitem[Goldreich(2020)]{G20}
Oded Goldreich.
\newblock The uniform distribution is complete with respect to testing identity
  to a fixed distribution.
\newblock In \emph{Computational Complexity and Property Testing: On the
  Interplay Between Randomness and Computation}, pages 152--172. Springer,
  2020.

\bibitem[Goldreich et~al.(1998)Goldreich, Goldwasser, and Ron]{GGR98}
Oded Goldreich, Shari Goldwasser, and Dana Ron.
\newblock Property testing and its connection to learning and approximation.
\newblock \emph{Journal of the ACM (JACM)}, 45:\penalty0 653--750, 1998.

\bibitem[Gutman(2002)]{G02}
Michael Gutman.
\newblock Asymptotically optimal classification for multiple tests with
  empirically observed statistics.
\newblock \emph{IEEE Transactions on Information Theory}, 35\penalty0
  (2):\penalty0 401--408, 2002.

\bibitem[Levi et~al.(2013)Levi, Ron, and Rubinfeld]{LRR13}
Reut Levi, Dana Ron, and Ronitt Rubinfeld.
\newblock Testing properties of collections of distributions.
\newblock \emph{Theory of Computing}, 9:\penalty0 295--347, 2013.

\bibitem[Mannor and Tsitsiklis(2004)]{MT04}
Shie Mannor and John~N Tsitsiklis.
\newblock The sample complexity of exploration in the multi-armed bandit
  problem.
\newblock \emph{Journal of Machine Learning Research}, 5\penalty0
  (Jun):\penalty0 623--648, 2004.

\bibitem[Marshall et~al.(2011)Marshall, Olkin, and Arnold]{MOA11}
Albert~W. Marshall, Ingram Olkin, and Barry~C. Arnold.
\newblock \emph{Inequalities: Theory of Majorization and its Applications}.
\newblock Springer, second edition, 2011.

\bibitem[Paninski(2008)]{P08}
Liam Paninski.
\newblock A coincidence-based test for uniformity given very sparsely sampled
  discrete data.
\newblock \emph{IEEE Transactions on Information Theory}, 2008.

\bibitem[Scheff{\'e}(1947)]{S47}
Henry Scheff{\'e}.
\newblock A useful convergence theorem for probability distributions.
\newblock \emph{The Annals of Mathematical Statistics}, 18, 1947.

\bibitem[Valiant and Valiant(2011)]{VV11}
Gregory Valiant and Paul Valiant.
\newblock The power of linear estimators.
\newblock In \emph{Symposium on Foundations of Computer Science (FOCS)}. IEEE,
  2011.

\bibitem[Yang et~al.(2024)Yang, Zhong, and Tan]{YZT24}
Junwen Yang, Zixin Zhong, and Vincent~YF Tan.
\newblock Optimal clustering with bandit feedback.
\newblock \emph{Journal of Machine Learning Research (JMLR)}, 2024.

\bibitem[Yavas et~al.(2025)Yavas, Huang, Tan, and Scarlett]{YHTS25}
Recep~Can Yavas, Yuqi Huang, Vincent~YF Tan, and Jonathan Scarlett.
\newblock A general framework for clustering and distribution matching with
  bandit feedback.
\newblock \emph{IEEE Transactions on Information Theory}, 2025.

\bibitem[Zhuang et~al.(2022)Zhuang, Chen, and Yang]{ZCY22}
Yubo Zhuang, Xiaohui Chen, and Yun Yang.
\newblock Wasserstein $ k $-means for clustering probability distributions.
\newblock \emph{Advances in Neural Information Processing Systems (NeurIPS)},
  35, 2022.

\bibitem[Ziv(2002)]{Z02}
Jacob Ziv.
\newblock On classification with empirically observed statistics and universal
  data compression.
\newblock \emph{IEEE Transactions on Information Theory}, 34\penalty0
  (2):\penalty0 278--286, 2002.

\end{thebibliography}
